\documentclass[superscriptaddress,nofootinbib,twocolumn,aps]{revtex4-2}
\usepackage[normalem]{ulem}
\usepackage{epsfig,amsmath,color,algorithm,algcompatible,amsthm,bm}
\usepackage{multirow}
\usepackage{mathbbol}
\usepackage{comment}
\usepackage[colorinlistoftodos,prependcaption]{todonotes}

\newcommand{\pro}[2]{\langle{#1}|{#2}\rangle}
\newcommand{\bra}[1]{\langle{#1}|}
\newcommand{\ket}[1]{|{#1}\rangle}
\newcommand{\bold}[1]{{\bf #1}}

\newcommand{\fig}[4]{
\begin{figure*}[ht]
\centering
\includegraphics[#4]{#1}
\caption{#2}
\label{#3}
\end{figure*}
}

\newtheorem{th.}{Theorem}
\newtheorem{co.}{Corollary}

\begin{document}

\title{
	Grover search revisited; application to image pattern matching
}

\author{Hiroyuki~Tezuka}
\affiliation{Sony Group Corporation, 1-7-1 Konan, Minato-ku, Tokyo, 108-0075, Japan}
\affiliation{Quantum Computing Center, Keio University, Hiyoshi 3-14-1, Kohoku-ku, Yokohama 223-8522, Japan}
\affiliation{Graduate School of Science and Technology, Keio University, 
Hiyoshi 3-14-1, Kohoku-ku, Yokohama, 223-8522, Japan}

\author{Kouhei~Nakaji}
\affiliation{Quantum Computing Center, Keio University, Hiyoshi 3-14-1, 
Kohoku-ku, Yokohama 223-8522, Japan}
\affiliation{Graduate School of Science and Technology, Keio University, 
Hiyoshi 3-14-1, Kohoku-ku, Yokohama, 223-8522, Japan}

\author{Takahiko~Satoh}
\affiliation{Quantum Computing Center, Keio University, Hiyoshi 3-14-1, 
Kohoku-ku, Yokohama 223-8522, Japan}
\affiliation{Graduate School of Science and Technology, Keio University, 
Hiyoshi 3-14-1, Kohoku-ku, Yokohama, 223-8522, Japan}

\author{Naoki~Yamamoto\thanks{
		e-mail address: \texttt{yamamoto@appi.keio.ac.jp}
	}}
\affiliation{Quantum Computing Center, Keio University, Hiyoshi 3-14-1, 
Kohoku-ku, Yokohama 223-8522, Japan}
\affiliation{Department of Applied Physics and Physico-Informatics, 
Keio University, Hiyoshi 3-14-1, Kohoku-ku, Yokohama, 223-8522, Japan}

\begin{abstract}
The landmark Grover algorithm for amplitude amplification serves as an essential 
subroutine in various type of quantum algorithms, with guaranteed quantum speedup 
in query complexity. 
However, there have been no proposal to realize the original motivating application 
of the algorithm, i.e., the database search or more broadly the pattern matching in 
a practical setting, mainly due to the technical difficulty in efficiently implementing 
the data loading and amplitude amplification processes. 
In this paper, we propose a quantum algorithm that approximately executes the entire 
Grover database search or pattern matching algorithm. 
The key idea is to use the recently proposed approximate amplitude encoding method on 
a shallow quantum circuit, together with the easily implementable inversion-test operation 
for realizing the projected quantum state having similarity to the query data, followed 
by the amplitude amplification operation that is independent to the target data index. 
We provide a thorough demonstration of the algorithm in the problem of image pattern 
matching. 
\end{abstract}


\maketitle

\section{Introduction}
\label{SEC_intro}

Grover search algorithm \cite{Grover1996-bh} is a landmark quantum algorithm 
that theoretically promises a computational advantage over any classical one. 
The search problem is included in the following pattern matching problem. 
That is, we are given a set of data, i.e., a database, where each data has 
its own index; the task is to find out the index of a database component that 
has the highest similarity to a given query data. 
When the database contains the query data and the task is to find the 
corresponding unique index, then the problem is called the search. 
In the original Grover's paper, he studied a simplified problem such that 
only the indices are focused, and provided the seminal amplitude amplification 
method that enables us to find the answer with $\sqrt{n}$ queries while any 
classical one needs $n$ queries, where $n$ is the number of data; that is, 
the quadratic speedup is guaranteed.

In fact, we find many studies for implementing the original Grover algorithm. 
However, to our best knowledge, those are not a practical one in the above 
sense; that is, there has been no proposal to implement the ``practical 
Grover algorithm” that gives the solution (index) for the search or pattern 
matching problem for a realistic database, with guaranteed quadratic speedup 
in the number of queries. 
This is because, in our view, there are two obstacles. 
The first issue is the difficulty to prepare the quantum states of database 
and query. 
In general, to load a data vector onto a $n$-qubit quantum state, $O(2^n)$ 
quantum gates are required 
\cite{Grover2000-xd,Sanders2019-rh,Plesch2011-io,Shende2006-qz}. 
That is, the number of gates required for the data loading increases 
exponentially with the number of qubits, which might destroy the quantum 
advantage. 
To address this issue, sometimes the Quantum Random Access Memory (QRAM) 
is assumed \cite{Giovannetti2008-ik,Giovannetti2008-sr}, from which an 
arbitrary quantum state is loaded, but realization of QRAM seems to be 
difficult. 
The second issue, which is though less serious than the first one, is that 
the operator for amplitude amplification often boils down to an ``Oracle" 
operator. 
This is clearly an obstacle for the practical use of Grover algorithm 
\cite{Gilliam2021-gi}, because Oracle is the operator constructed with the 
answer of the problem; hence many previous studies treat Oracle as a 
black-box function \cite{Liu2021-fk,Kasirajan2021-va}.

In this paper, we propose a coherent method for realizing the Grover algorithm 
for search or pattern matching problems, which circumvents the above-mentioned 
two issues. 
As for the first issue, we employ the method of {\it approximate amplitude 
encoding (AAE)} that uses a constant-depth parameterized quantum circuit (PQC) 
for the data loading onto a quantum state \cite{Nakaji2021-as}; 
in our case, we use AAE to prepare both the database state and the query 
state. 
Secondly, we formulate the amplitude amplification process such that the 
operators in the algorithm can be implemented without using the oracle, 
or in other words without knowing the index of query state (i.e., the answer). 
The key of our algorithm is to use the so-called inversion-test technique 
to realize projection of the database state onto a subspace of 
states that have overlap with the query state; the amplitude of components 
of the projected state can be then amplified by the operator that does 
not contain the answer index. 
This framework may also be applicable to recently proposed shallow Grover 
algorithms \cite{Liu2021-fk,Brianski2021-vd}, which make the diffusion 
operator shallower and thus preferable in noisy intermediate-scale quantum 
(NISQ) \cite{preskill2018quantum} devices. 
Of course, it is also beneficial for future fault-tolerant quantum 
computers (FTQC).

We demonstrate our algorithm in the framework of quantum image processing 
(QIMP) \cite{Ruan2021-qs} in numerical simulations; particularly the error 
analysis of the circuit without amplitude amplification is experimentally 
conducted using the IBM superconducting quantum device. 
In QIMP, by embedding an image information onto a quantum state, we can 
process various tasks efficiently with much less number of bits and 
queues compared to the classical case. 
The pattern matching, which has various applications such as real-time 
object recognition, is one such task \cite{Jiang2016-ji,Zhou2018-xj,Liu2019-bb,Guanlei2020-em,Iliyasu2016-wo,Dang2017-te}. 
However, they did not discuss the issue of state preparation, rather 
assumed that both the database and query image data are perfectly loaded 
onto quantum states. 
Moreover, most of those works employ an inefficient classical strategy 
that compares the query quantum state with each data quantum state in 
the database one by one. 
Clearly, we may develop the quantum strategy that prepares a single 
quantum state representing the whole database and further uses Grover 
algorithm for amplifying the hitting probability. 
In fact, Ref. \cite{Zhou2018-xj} employs the amplitude amplification 
operation, which is though used to only enhance the above-mentioned 
classical strategy. 
Our proposed method resolves all these issues; both the query and 
database quantum states are prepared via AAE, and Grover operator 
is constructed without knowing the target indices, for realizing the 
entire quantum image pattern matching algorithm applicable for realistic 
image dataset. 
In this sense, the demonstration itself is a contribution to the 
area of QIMP.

The rest of the paper is organized as follows. 
In Section~\ref{SEC_algo}, we describe our enhanced pattern matching 
algorithm composed of the amplitude amplification (\ref{SEC_algo_aa}) 
and AAE (\ref{SEC_algo_aae}). 
Section~\ref{SEC_demo} demonstrates the application of the algorithm to 
the image pattern matching problem, via the numerical simulation and 
the experiment with real quantum device. 
Section~\ref{SEC_concl} concludes this paper and discuss possible 
directions of future research.


\section{Quantum pattern matching algorithm}
\label{SEC_algo}

\subsection{Basic algorithm}
\label{SEC_algo_qpm}

\subsubsection*{Problem setting}

Our problem is described as follows. 
First, we have a {\it database} composed of $N_D$ dimensional real data 
vectors ${\bf a}_k = [a_{0,k}, \ldots, a_{N_D-1, k}]\mbox{}^\top$, 
where $k=0, \ldots, N_I-1$ denotes the index of data; 
that is, the database contains $N_I$ data vectors. 
Then we are given a {\it query} data 
${\bf b} = [b_0, \ldots, b_{N_D-1}]\mbox{}^\top$, 
which is also a $N_D$ dimensional real vector. 
They will be encoded into quantum states and for this reason assumed to 
be normalized, i.e., $\sum_j a_{jk}^2=1$ and $\sum_j b_{j}^2=1$. 
Our goal is to identify the index, which we call the {\it target index}, of 
the database components that has the largest overlap (similarity) to the 
query ${\bf b}$; 
that is, $k_* = {\rm argmax}_k |{\bf a}_k^\top {\bf b}|$.

\subsubsection*{Step 1: Preparation of database and query quantum states}
\label{SEC_algo_step1}

To execute the above-mentioned pattern matching task on a quantum device, 
we first need to load the data onto quantum states. 
That is, we prepare the database state $|{\rm database}\rangle$ and the query state 
$\ket{\rm query}$ as follows: 
\begin{align}
	\label{EQ_step1_psi_1}
	|{\rm database} \rangle & = A |0\rangle^{\otimes (n_D+n_I)} \\
	&= \frac{1}{\sqrt{N_{I}}} \sum_{j=0}^{N_D -1} \sum_{k=0}^{N_I -1} a_{jk} |j\rangle_{D} |k\rangle_{I}  \\
	\label{EQ_step1_psi_data_index}
	&\equiv \frac{1}{\sqrt{N_{I}}} \sum_{k}|{\rm data}(k)\rangle \otimes |k\rangle_{I}, \\
	\label{EQ_step1_phi_1}
	\ket{\rm query} & = B |0\rangle^{\otimes n_D} \\
	&= \sum_{j=0}^{N_D -1} b_{j} |j\rangle_{D}. 
\end{align}
Here $\{\ket{j}_D\}$ and $\{\ket{k}_I\}$ are the orthogonal computational 
basis set in the data Hilbert space ${\cal H}_D$ and the index Hilbert 
space ${\cal H}_I$, respectively. 
For simplicity, we assume that these spaces are identical to those of 
$n_D$ and $n_I$ qubits, meaning that $N_D=2^{n_D}$ and $N_I=2^{n_I}$. 
In Eq.~\eqref{EQ_step1_psi_data_index}, we define the quantum state 
corresponding to the $k$th data vector 
$|{\rm data}(k)\rangle = \sum_{j=0}^{N_D -1} a_{jk} |j\rangle_{D}$; 
that is, the database state $|{\rm database}\rangle$ is realized as a 
superposition of all data vectors accompanied with their indices. 
The database operator $A$ and the query operator $B$ are unitary operators 
for generating the corresponding quantum states. 
Note that, in the above expression, $A$ and $B$ are assumed to realize 
the perfect data encoding, in which case, however, exponential number of 
gate operations have to be contained in the corresponding quantum circuits 
\cite{Grover2000-xd,Sanders2019-rh,Plesch2011-io,Shende2006-qz}. 
In Section \ref{SEC_algo_aae}, we will introduce the constant-depth circuits 
that approximate $A$ and $B$.

Before moving to the next step, we remark that, instead of the above amplitude 
encoding, we can use the following basis encoded state: 
\begin{align}
\label{basis encode II-A database}
	|{\rm database}\rangle 
	&= \frac{1}{\sqrt{N_{I}}} \sum_{j=0}^{N_D -1} \sum_{k=0}^{N_I -1} \hat{a}_{jk} |j\rangle_{D} |k\rangle_{I},  \\
\label{basis encode II-A query}
    \ket{\rm query} 
	&= \frac{1}{\sqrt{N_{BE}}} \sum_{j=0}^{N_D -1} \hat{b}_j |j\rangle_{D}. 
\end{align}
Here $\hat{a}_{jk}$ and $\hat{b}_j$ are binary (i.e., 0 or 1) variables that are determined 
from the original information $a_{jk}$ and $b_j$. 
Also ${\bf \hat{a}}$ has $N_I$ non-zero components, and $N_{BE}$ is the normalization constant for the query data encoded in basis encoding. 
For instance, under $n_{I}=n_{D}=2$, ${\bf \hat{b}} = [\hat{b}_{0}, \hat{b}_{1}, \hat{b}_{2}, \hat{b}_{3}]\mbox{}^\top = [0, 1, 0, 0]\mbox{}^\top$ if the query data is $\ket{1}_D$, 
and ${\bf \hat{a}} = [\hat{a}_{00}, \hat{a}_{10}, \hat{a}_{20}, \hat{a}_{30}, \cdots , \hat{a}_{23}, \hat{a}_{33}]\mbox{}^\top = [0, 1, 0, 0, \cdots , 0, 0]\mbox{}^\top$ if the database contains $\ket{1}_D$ at the index $\ket{0}_I$.

\subsubsection*{Step 2: Basic algorithm for computing the similarity}
\label{SEC_algo_step2}

The inner product of the classical data vectors ${\bf a}_k$ and ${\bf b}$ 
is now, in terms of quantum states, represented as the fidelity 
\begin{equation*}
   \pro{\rm query}{{\rm data}(k)} 
      = \bra{0^{n_D}} B^\dagger \ket{{\rm data}(k)}, 
\end{equation*}
where we used the simplified notation $\ket{0^n}=\ket{0}^{\otimes n}$. 
This can be represented as the fidelity between 
\begin{equation}
    \ket{{\rm query}'} =\ket{0}^{\otimes n}
\end{equation}
and $B^\dagger \ket{{\rm data}(k)}$. 
Our algorithm uses the inversion test technique to evaluate this quantity; 
that is, naively, the fidelity can be computed as the probability to obtain 
$\ket{{\rm query}'}$ when measuring the state $B^\dagger \ket{{\rm data}(k)}$ 
in the computational basis.

\begin{figure}[t!] 
   \includegraphics[width=3.3in]{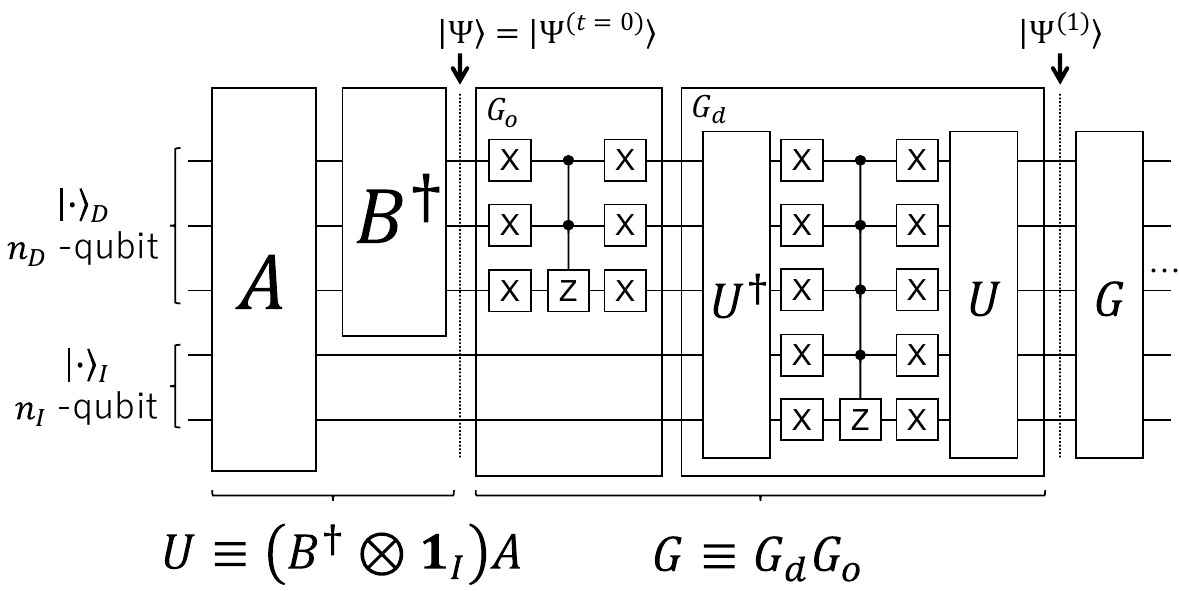}
   \caption{The entire quantum circuit for pattern matching, composed of the 
   AAE circuit $A$ and $B$, followed by the Grover circuit $G$.}
   \label{FIG_circuit_0}
\end{figure}

Now, each $\ket{{\rm data}(k)}$ is not directly given to us, but rather we have 
$\ket{{\rm database}}=A\ket{0}^{\otimes (n_D+n_I)}$. 
Hence the state for inversion test is given by 
\begin{align*}
	|\Psi\rangle & \equiv (B^{\dag}\otimes \mathbb{1}_{I}) 
	A\ket{0}^{\otimes (n_D+n_I)} \\
	&= \frac{1}{\sqrt{N_{I}}} \sum_{k}B^{\dag}|{\rm data}(k)\rangle \otimes |k\rangle_I, 
\end{align*}
where $\mathbb{1}_{I}$ is the identity operator on ${\cal H}_I$. 
Now we make the computational-basis measurement on the first 
$n_D$ qubits of $|\Psi\rangle$ and post-select the state when the result is all zeros. 
The entire quantum circuit for executing this task is illustrated in Fig.~\ref{FIG_circuit_0} 
($G$ is the Grover operator described later). 
The resultant state is given by 
\begin{equation}
\label{EQ_result_postselect}
    |R\rangle = \frac{1}{C_R} \sum_{k} r_{k} |k\rangle_I,
\end{equation}
where 
\begin{align*}
r_{k} = \langle {\rm query} | {\rm data}(k)\rangle, ~~
C_R = \sqrt{\sum_{k} |\langle {\rm query} | {\rm data}(k)\rangle|^2}.
\end{align*}
Thus, by further measuring the state \eqref{EQ_result_postselect} in the 
computational basis of ${\cal H}_I$, we are likely to obtain the index $k$ with 
relatively large probability proportional to 
$|\langle {\rm query}| {\rm data}(k)\rangle|^2$, which is indeed the goal of the 
pattern matching task.

However, the above approach has a critical issue that the success probability of 
the post-selection and as a result the probability of hitting the target index 
are severely suppressed, when the number of database components $N_I$ is large. 
More precisely, the probability that the post-selection is succeeded and 
subsequently the target index $k_*$ is identified, is calculated as
\begin{align*}
	P({\rm index}=k_*) 
	&= | \mbox{}_I\langle k_*| \langle 0^{n_D} |\Psi\rangle|^2 \\
	&= \frac{1}{N_I} |\langle {\rm query}|{\rm data}(k_*)\rangle |^2 \\
	&\le \frac{1}{N_I}. 
\end{align*}
This means that we need $O(N_I)$ measurements for identifying the target index, 
which is the same computational complexity as that of classical algorithms. 
This is indeed the motivation to introduce the Grover algorithm; 
that is, we may use the Grover algorithm to amplify the amplitude of 
$\ket{0}^{\otimes n_D}\ket{k}_I$ in the state $\ket{\Psi}$ for any $k$. 
The next subsection is devoted to describe the scheme in detail.


\subsection{Amplitude amplification}
\label{SEC_algo_aa}

First, we explain the point of the algorithm. 
The purpose of amplitude amplification is to amplify the amplitude of 
state components of $|{\rm database}\rangle=A|0\rangle ^{\otimes n_D+n_I}$, which 
have an overlap with $|{\rm query}\rangle=B|0\rangle ^{\otimes n_D}$. 
This is equivalent to amplifying the amplitude of states of 
$|\Psi \rangle=(B^{\dag}\otimes 
\mathbb{1})A|0\rangle ^{\otimes n_D+n_I}$ that have an overlap with 
$|{\rm query}^{\prime}\rangle=|0\rangle ^{\otimes n_D}$; 
this simple trick brings a big benefit in view of the implementation that, 
as shown later, the projection part (the oracle part) contained in the 
Grover operator is simply that on the computational basis 
$|0\rangle ^{\otimes n_D}$ rather than that on the entangled state 
$|{\rm query}\rangle$. 
This is called the inversion test technique, which can be now employed 
thanks to the explicit construction of the encoding process $B$; note 
also that the resulting Grover operator is not anymore an oracle, but 
is what can be explicitly implementable without knowing the target index. 
Once the Grover operator is constructed, it efficiently amplifies the 
amplitudes that are proportional to the similarity between the 
database component and the query, which corresponds to all non-zero 
$r_k$ in Eq.~\eqref{EQ_result_postselect}. 
This is an extended framework of the conventional Grover algorithm; that is, the 
target states are distributed \cite{Ezhov2000-sf}, rather than given as one of the 
basis state of the initial state. 
Below we apply this theory to our problem; see Appendix~\ref{SEC_grover_dq} for the 
detailed calculation.

First, the initial state $|\Psi\rangle$ and the target state $|q\rangle$ are defined as 
\begin{align}
	\label{EQ_def_psi'}
	|\Psi\rangle 
	   & = (B^{\dag}\otimes \mathbb{1}_{I}) A\ket{0}^{\otimes (n_D+n_I)}, \\
	|q\rangle & \equiv 
	\frac{\left( |0^{n_D}\rangle \langle 0^{n_D}| \otimes \mathbb{1}_I \right) 
	             |\Psi \rangle}
	       {\sqrt{\langle \Psi | \left( |0^{n_D}\rangle \langle 0^{n_D}| \otimes \mathbb{1}_I \right)
	             |\Psi \rangle }}. 
\end{align}
Recall the notation $\ket{0^n}=\ket{0}^{\otimes n}$. 
That is, the target state $\ket{q}$ is the component of $\ket{\Psi}$ that has an overlap with 
$|{\rm query}^{\prime}\rangle=|0\rangle ^{\otimes n_D}$ in the data Hilbert space ${\cal H}_D$. 
More specifically, $\ket{\Psi}$ can be expressed as 
\begin{equation}
     \ket{\Psi} = \sin\theta \ket{q} + \cos\theta \ket{q^\perp}, 
\end{equation}
where $\ket{q^\perp}$ is the state orthogonal to $\ket{q}$, and $\sin\theta = \pro{q}{\Psi}$. 
Also, because the state in our scenario is distributed in the basis states, it is convenient 
to have the expression of the above states in terms of the basis vectors as follows: 
\begin{align}
\label{basis states expansion}
    |\Psi\rangle = \sum_{x=0}^{N-1} \psi_x |x\rangle, ~~~
    |q\rangle = \sum_{x=0}^{N-1} q_x |x\rangle,
\end{align}
where $N=N_D N_I$ is the dimension of the entire Hilbert space 
${\cal H}_D \otimes {\cal H}_I$, and $\{ |x\rangle \}$ is the set of basis states 
of this space. 
Then $\{b_x\}$ has at most $ N_{I}$ non-zero components, which can be 
explicitly represented as follows; 
\begin{equation*}
	q_x =
	\begin{cases}
	    \psi_{x}/\langle q | \psi\rangle  & \mathrm{if} \; x \in {\cal C}, \\
	    0   & \mathrm{if} \; x \notin {\cal C}, 
	    \end{cases}
\end{equation*}
where ${\cal C}$ is the set of numbers defined as 
\begin{equation*}
     {\cal C} = \{ x ~|~ \bra{x} (\ket{0^{n_D}}\bra{0^{n_D}}\otimes \mathbb{1}_I) \ket{x} \neq 0 \}. 
\end{equation*}
Note that $\pro{q}{\Psi} 
= \left[ \bra{\Psi} (|0^{n_D}\rangle \langle 0^{n_D}| \otimes \mathbb{1}_I ) \ket{\Psi} \right]^{1/2}$.

The goal of amplitude amplification is to amplify the coefficient $\sin\theta$ via applying the Grover operator, which is 
decomposed of the following oracle operator $G_o$ and the diffusion operator $G_d$: 
\begin{align*}
      |\Psi\mbox{}^{(t+1/2)}\rangle 
        &= G_{o}|\Psi\mbox{}^{(t)}\rangle \\ 
        &= (\mathbb{1} -2|0^{n_D}\rangle \langle 0^{n_D}| \otimes \mathbb{1}_I)
                |\Psi\mbox{}^{(t)}\rangle, \\
      |\Psi\mbox{}^{(t+1)}\rangle &= G_{d}|\Psi\mbox{}^{(t+1/2)}\rangle \\
        &= U(2|0^{n_D+n_I}\rangle \langle 0^{n_D+n_I}| -\mathbb{1}_{DI}) U^{\dag} 
                |\Psi\mbox{}^{(t+1/2)}\rangle \\
        &= (2|\Psi\rangle \langle \Psi| -\mathbb{1}_{DI})
	       |\Psi\mbox{}^{(t+1/2)}\rangle, 
\end{align*}
where $U=(B^{\dag}\otimes \mathbb{1}_{I})A$ is the unitary operator producing 
the initial state, i.e., $|\Psi^{(0)}\rangle = |\Psi\rangle = U |0\rangle ^{\otimes n_{D}+n_{I}}$. 
That is, the amplified state $|\Psi^{(t)}\rangle$ is generated by $t$ applications of the Grover 
operator $G=G_d G_o$ on the initial state $|\Psi\rangle$. 
The oracle operator $G_{o}$ flips the phase of the target state $\ket{q}$, 
while it does not change any state orthogonal to $\ket{q}$. 
The diffusion operator $G_{d}$ inverts the amplitudes around their averaged value. 
Note that, although $G_o$ is called the oracle by convention, it can be composed 
without any information about the target index (the answer). 
The entire circuit composed of the encoding and amplitude amplification operators is depicted in 
Fig.~\ref{FIG_circuit_0}.

Following the general theory given in Appendix~\ref{SEC_grover_dq}, the 
state after $t$ Grover operations is explicitly calculated as 
\begin{equation*}
    |\Psi ^{(t)} \rangle 
       = G^{t} |\Psi \rangle
       = \sin((2t+1)\theta) |q \rangle + \cos((2t+1)\theta) |q^{\perp} \rangle, 
\end{equation*}
showing that the amplitude of $\ket{q}$ can be amplified by appropriately choosing 
the number of operations. 
Also, corresponding to Eq.~\eqref{basis states expansion}, the transformed state is expressed as 
\begin{equation*}
	|\Psi^{(t)}\rangle = \sum_{x} \psi_{x}^{(t)} |x\rangle, ~~~
	\psi_{x}^{(t)} = A_{x} \sin{(\omega t +\delta_{x})}, 
\end{equation*}
where $\omega$, $A_{x}$, and $\delta_{x}$ (i.e., the frequency, the amplitude, 
and the phase of the initial state, respectively) are given by 
\begin{equation}
    \begin{aligned}
    \label{EQ_omega}
    \omega &= 2 \arcsin{\langle q|\Psi\rangle}, 
    \end{aligned}
\end{equation}
\begin{equation}
    \begin{aligned}
	\label{EQ_Ax}
	A_x &= \sqrt{\frac{q_{x}^2 -2\langle q| \Psi \rangle \psi_{x} q_{x} +\psi_{x}^{2}}
	                            {1-\langle q| \Psi \rangle^2}} \\
	&=
	    \begin{cases}
	        \psi_{x}/\langle q | \Psi\rangle               & \mathrm{if} \; x \in {\cal C}, \\
	        \psi_{x}/\sqrt{1- \langle q | \Psi\rangle^2}   & \mathrm{if} \; x \notin {\cal C},
	    \end{cases}
	\end{aligned}
\end{equation}
\begin{equation}
    \begin{aligned}
	\label{EQ_deltax}
	\delta_x &= \arccos \left( \frac{q_{x}-\langle q| \Psi\rangle \psi_{x}}
	{\sqrt{q_{x}^2 -2\langle q| \Psi\rangle \psi_{x} q_{x} +\psi_{x}^{2}}} \right) \\
	&=
	    \begin{cases}
	        \arccos{\sqrt{1-\langle q | \Psi\rangle ^2}} & \mathrm{if} \; 
	        x \in {\cal C}, \\
	        \arccos{(-\langle q | \Psi\rangle)} & \mathrm{if} \; 
	        x \notin {\cal C},
	    \end{cases}
	\end{aligned}
\end{equation}
where $\langle q| \Psi\rangle 
= \left[ \langle \Psi| (|0^{n_D}\rangle \langle 0^{n_D}| \otimes \mathbb{1}_I ) |\Psi \rangle \right]^{1/2}$. 
Note that the phases of $\psi_x^{(t)}$ for the case $x\in{\cal C}$ and that for the case $x\notin{\cal C}$ 
are shifted just $\pi /2$ with each other.

Based on these results, we obtain the analytic expression of the amplified 
probability of hitting the matched indices; that is, for the index $x\in{\cal C}$, 
the hitting probability after $t$ Grover iterations is given by 
\begin{align}
    P_{x}^{(t)} &= \{ \psi_{x}^{(t)} \} ^2 \nonumber \\
    \label{EQ_Px_AA_2}
    &= \frac{\psi_{x}^{2}}{2 \langle q| \Psi\rangle ^2} 
            \left[ 1-\cos{2{ \left(\omega t +\arccos{\sqrt{1-\langle q | \Psi\rangle ^2}} \right)}} \right] \\
    &\leq \frac{\psi_{x}^{2}}{\langle q| \Psi\rangle ^2}. \nonumber
\end{align}
From Eq.~\eqref{EQ_Px_AA_2}, we can determine the optimal number of iteration 
$t_*$ that gives the highest hitting probability $P_{x}$ for any index $x\in{\cal C}$ 
as follows: 
\begin{align}
    \label{EQ_Px_CI}
    t_{*} = CI \left( \frac{\arccos\langle q| \Psi \rangle}
                        {2 \arcsin \langle q| \Psi \rangle} \right),
\end{align}
where $CI(z)$ returns the closest integer of a real number z by rounding down.

The above result \eqref{EQ_Px_CI} shows that the optimal operations number $t_*$ 
depends on the initial overlap 
$\langle q| \Psi\rangle 
= \left[ \langle \Psi| (|0^{n_D}\rangle \langle 0^{n_D}| \otimes \mathbb{1}_I ) |\Psi \rangle \right]^{1/2}$. 
That is, for the precise treatment of the problem, we need to estimate 
$\langle q| \Psi\rangle$. 
As indicated by this expression, this task can be conducted by estimating 
the success probability of projecting $\ket{\Psi}$ onto $\ket{0}^{\otimes N_D}$; 
but this strategy is inefficient in the sense that it needs to prepare $\ket{\Psi}$ for 
$O(1/\epsilon^2)$ times, where $\epsilon$ denotes the given estimation error. 
Instead, we could take the sophisticated amplitude estimation method 
\cite{brassard2002quantum,Nakaji2020-sa,suzuki2020amplitude}. 
With the use of this technique, the parameter $\omega$ and accordingly 
$\langle q| \Psi\rangle$ can be estimated via $O(1/\epsilon)$ operations of Grover. 
However, in a practical case where the number of database, $N_I$, is enough 
large and the database contains only few data similar to the query, we have a rough 
estimate $\omega \simeq \langle q|\Psi \rangle \simeq 1/\sqrt{N_{I}}$.
In this case we obtain $t_{*} \simeq \sqrt{N_{I}}$ and further 
$P_{x}^{(t_{*})} \simeq N_{I}\psi_{x}^{2} \simeq 1$ if $\psi_{x}^{2} \rightarrow 1/{N_{I}}$. 
This clearly means the quadratic speedup in the number of operations necessary to 
achieve the pattern matching, similar to the conventional Grover search algorithm.

Lastly, we remark on the implementation of the circuit. 
In our framework, the oracle operator $G_o$ is simple, and it can be composed of 
one $n_D$-controlled Toffoli gate. 
In contrast, the diffusion operator $G_d$ contains $U^\dag$ and $U$, which consist 
of the database operator $A$ and the query operator $B$. 
If we naively implement the perfect version of those data loading operators, the number 
of gates increases exponentially in the system size, i.e., $O(2^{n_D +n_I})$. 
This is severe to implement on near future devices, and might spoil the quantum 
advantages even on fault-tolerant quantum devices. 
To overcome this difficulty, we introduce AAE, as described in the next subsection.


\subsection{Approximate amplitude encoding (AAE)} 
\label{SEC_algo_aae}

\fig{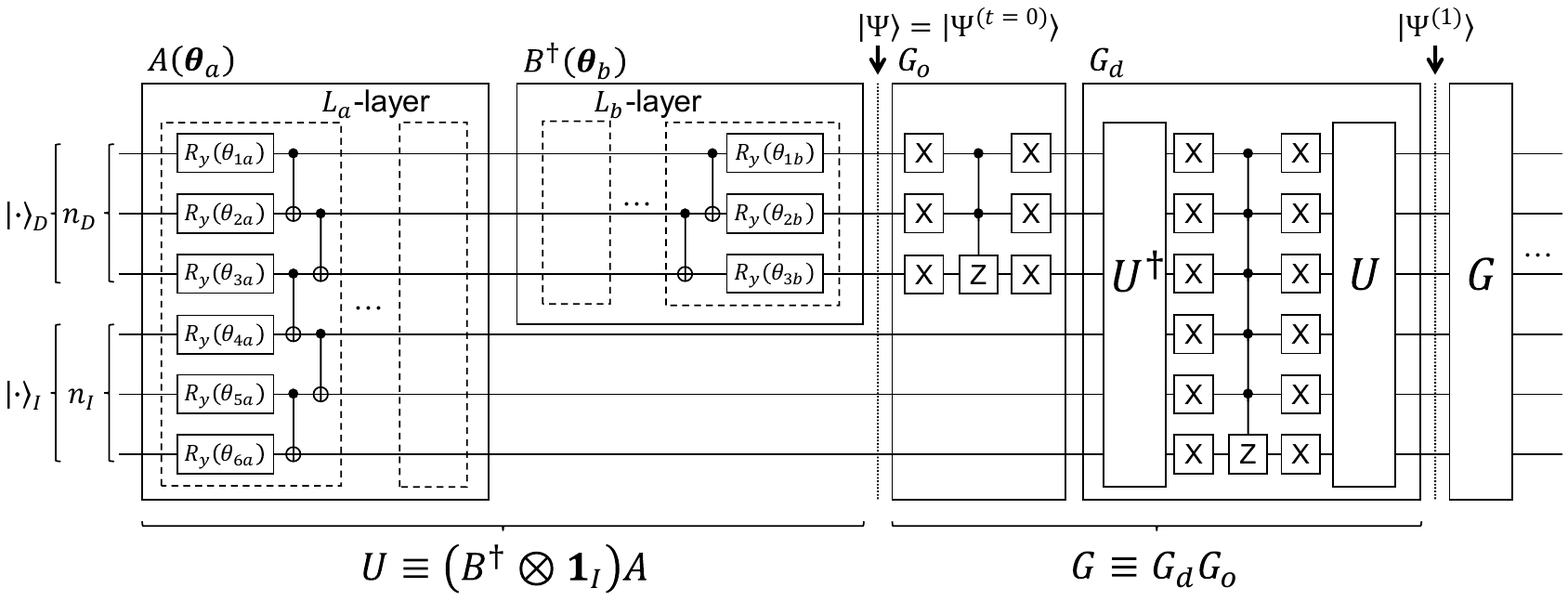}
{A detailed schematic of the entire quantum circuit for pattern matching, 
in the case $(n_{D},n_{I})=(3,3)$.}
{circuit}{width=\textwidth}

AAE \cite{Nakaji2021-as} is an algorithm that trains a PQC that realizes approximate 
data loading. 
Given a target $n$-qubit state with real amplitudes as $|\boldsymbol{d}\rangle$, a PQC 
$U(\boldsymbol{\theta})$ with parameters $\boldsymbol{\theta}$ is trained so that 
$U(\boldsymbol{\theta})|0\rangle^{\otimes n}$ approximates $e^{i\alpha}|\boldsymbol{d}\rangle$ 
where $e^{i\alpha}$ is the global phase. 
AAE runs different algorithms depending on two cases: Case~1 and Case~2. 
Case~1 is the case where the amplitudes of target quantum state represented in the computational 
basis are all non-negative or non-positive. 
Otherwise (i.e., for Case~2), AAE offers a different algorithm. 
The numerical experiment shown in Section~\ref{SEC_demo} corresponds to Case 1, 
and therefore, we review only Case~1 here.

The goal of AAE is to find $U(\boldsymbol{\theta})$ that ideally satisfies
\begin{equation}
    \label{EQ_AAE_1}
    e^{i\alpha}U(\boldsymbol{\theta})|0\rangle^{\otimes n} 
       = |\boldsymbol{d}\rangle \equiv \sum_{j=0}^{N-1} d_{j}|j\rangle,
\end{equation}
where $N=2^n$ and $d_j$ is the $j$th element of the $N$-element normalized real vector 
$\boldsymbol{d}$. 
To formulate a problem of finding such $U(\boldsymbol{\theta})$, we need a simple yet 
equivalent condition to Eq.~\eqref{EQ_AAE_1}; the important point is that each element of 
$e^{i\alpha}U(\boldsymbol{\theta})|0\rangle^{\otimes n}$ have to identify not only the absolute 
value but also the sign of  $d_{j}$. 
Actually Ref.~\cite{Zoufal2019-to} proposed the method to (approximately) load only the 
absolute value of the coefficients by utilizing the generative adversarial network (GAN).
In contrast, it was shown in \cite{Nakaji2021-as} that Eq.~\eqref{EQ_AAE_1} is equivalent 
to the following conditions: 
\begin{align}
	\label{EQ_caseone_condition_1}
	|\langle j|U(\boldsymbol{\theta})|0\rangle^{\otimes n}|^2 & = d_j^2, 
	\\
	\label{EQ_caseone_condition_2}
	|\langle j|H^{\otimes n}U(\boldsymbol{\theta})|0\rangle^{\otimes n}|^2
	                     & = \left(\sum_{k=0}^{N-1}d_k\langle j|H^{\otimes n}|k\rangle\right)^2 \\
	                     & \equiv \left(d_j^{H}\right)^2. 
	\nonumber
\end{align}
Note that $d_j^{H}$ is classically computable from $|{\rm data}\rangle$ with complexity 
$O(N\log N)$, by using the Walsh-Hadamard transform \cite{Ahmed1975-un}.

The training is performed so that the following cost function $\mathcal{L}$ is minimized 
by utilizing the gradient descendant algorithm: 
\begin{equation}
\label{AAE cost}
    \mathcal{L} = \frac{\mathcal{L}_1 + \mathcal{L}_2}{2}, 
\end{equation}
where 
\begin{equation*}
\begin{split}
    \mathcal{L}_1 &= \mathcal{L}_{MMD}
             (\{|\langle j|U(\boldsymbol{\theta})|0\rangle|^2\},\{d_j^2\}),  \\
    \mathcal{L}_2 &= \mathcal{L}_{MMD}
             (\{|\langle j|H^{\otimes n}U(\boldsymbol{\theta})|0\rangle|^2\},\{(d_j^H)^{2}\}). 
\end{split}
\end{equation*}
Here, $\mathcal{L}_{MMD}(\{q(j)\}, \{p(j)\})$ is the {\it maximum mean discrepancy} (MMD) 
between two discrete probability distributions $q(j)$ and $p(j)$ 
\cite{liu2018differentiable,coyle2020born}: 
\begin{equation*}
	\label{EQUATION-mmd-definition}
	\begin{split}
		\mathcal{L}_{MMD}(\{q(j)\}, \{p(j)\}) 
		         &\equiv \gamma_{ MMD}(\{q(j)\}, \{p(j)\})^2, \\
		\gamma_{MMD}(\{q(j)\}, \{p(j)\}) 
		         &= \left|\sum_{j=0}^{N-1} q(j)\bold{\Phi}(j) - \sum_{j=0}^{N-1} p(j)\bold{\Phi}(j)\right|,
	\end{split}
\end{equation*}
where $\bold{\Phi}(j)$ is a function that maps the discrete random variable $j$ 
to a feature space. 
It is shown that, as long as we choose $\bold{\Phi}(k)$ so that the kernel function 
$\kappa(j, k) \equiv \bold{\Phi}(j)\bold{\Phi}(k)$ is a Gaussian kernel, the 
condition $\mathcal{L}_{MMD}(\{q(j)\}, \{p(j)\})=0$ is equivalent to $q(j) = p(j)$ 
for all $j$. 
In this paper, we choose Gaussian kernel functions in both $\mathcal{L}_1$ and 
$\mathcal{L}_2$; hence the condition $\mathcal{L} = 0$ is equivalent to 
\eqref{EQ_caseone_condition_1} and \eqref{EQ_caseone_condition_2}, which 
implies the validity of minimizing the cost function $\mathcal{L}$.

By using AAE, the database state $|{\rm database}\rangle$ and the query state 
$|{\rm query}\rangle$ are approximately generated as follows; 
\begin{align}
    \label{EQ_tilde_a}
    |{\rm database}\rangle 
        & \simeq A(\boldsymbol{\theta}_{a}) |0\rangle^{\otimes n_D+n_I} \\
        &= \frac{1}{\sqrt{N_{I}}} \sum_{j=0}^{N_D -1} 
                \sum_{k=0}^{N_I -1} \tilde{a}_{jk} |j\rangle_{D} |k\rangle_{I}  \\
	    &= \frac{1}{\sqrt{N_{I}}} \sum_{k}|{\rm data}(k)\rangle \otimes |k\rangle_{I}, \\
	\label{EQ_tilde_b}
	|{\rm query}\rangle & \simeq B(\boldsymbol{\theta}_{b}) |0\rangle^{\otimes n_D} \\
	    &= \sum_{j=0}^{N_D -1} \tilde{b}_j |j\rangle_{D}. 
\end{align}
Here, $A(\boldsymbol{\theta}_a)$ and $B(\boldsymbol{\theta}_b)$ are parametrized 
unitary operators for generating the database state and the query state, respectively; 
$\boldsymbol{\theta}_a$ and $\boldsymbol{\theta}_b$ are the optimal parameter vectors 
that minimize the cost function \eqref{AAE cost} for each cases. 
Also, $\tilde{a}_{jk}$ and $\tilde{b}_j$ are the coefficient of the generated 
states by PQCs $A$ and $B$, respectively. 
The notation $\simeq$ in \eqref{EQ_tilde_a} and \eqref{EQ_tilde_b} represents 
the approximate encoding, meaning that some error may be contained.
These AAE operators are followed by the Grover operator, as shown in 
Fig.~\ref{circuit}, where a detailed gate structure of the AAE part is depicted.

Lastly note that AAE can be applied to the basis encoded state 
\eqref{basis encode II-A database} and \eqref{basis encode II-A query} as well as the above amplitude encoded one. 
One may employ the exact basis encoding method using Toffoli gates, because 
the construction is logically straightforward and actually some efficient 
encoding schemes have been proposed, such as \cite{Bergholm2005-pd}. 
However, the total number of gates may be drastically reduced by finding an 
approximating encoding circuit with the use of AAE, although the learning 
process of the parameters will bring some encoding error. 
One may choose the encoding technique, the exact encoding or AAE, depending on 
the situation.


\section{Application to image pattern matching}
\label{SEC_demo}

In this section, we provide a thorough numerical demonstration of our algorithm 
applied to the image pattern matching problem. 
A set of toy image data is considered, with and without the amplitude amplification. 
In particular, the error analysis of the circuit for AAE and the inversion-test is  
experimentally conducted using the IBM superconducting quantum device.


\subsection{General quantum states for database and query image data}

An image data consists of $N_P$ pixels, where each pixel has respective 
color intensity represented by the integer in the range of $[0, N_C-1]$. 
This means that the data vector is of $N_D = N_P N_C$ dimension. 
To encode this image data onto a quantum state, we take the basis encoding 
representation called Novel Enhanced Quantum Representation (NEQR); see 
Appendix~B for the detailed description. 
Note that, as explained in Section~\ref{SEC_algo_aae}, the basis encoding 
scheme can be handled via AAE.

\begin{figure}
  \includegraphics[width=3.3in]{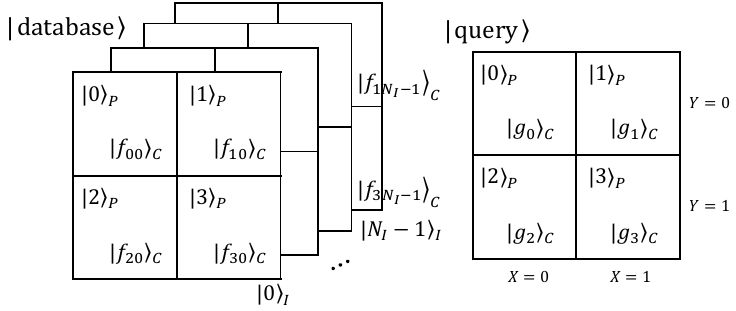}
  \caption{The database state (left) and the query state (right).}
  \label{FIG_fig_setting}       
\end{figure}

First, we take $n_C$ and $n_P$ qubits to represent the variable of color 
intensity and the pixel position, i.e., $N_C=2^{n_C}$ and $N_P=2^{n_P}$; 
then the total number of qubits of the quantum state corresponding to this 
image data is $n_D = n_C + n_P$. 
Now, let $g_j$ be the binary-represented integer that denotes the color 
intensity at the $j$th pixel of the query data; then in the NEQR format 
the corresponding quantum state is assigned as 
\begin{equation}
   \nonumber
   |{\rm query}\rangle 
   = \frac{1}{\sqrt{N_{P}}} 
        \sum_{j=0}^{N_P -1} |g_{j}\rangle_{C} |j\rangle_{P}. 
\end{equation}
Note that this state can be expressed in the form 
\eqref{basis encode II-A query} where $\hat{b}_j$ takes 1 only 
when $\ket{j}_D$ in Eq.~\eqref{basis encode II-A query} coincides with 
the above $|g_{j}\rangle_{C} |j\rangle_{P}$. 
Figure~\ref{FIG_fig_setting} depicts the case where $N_P = 4$. 
The database quantum state is constructed in the same way. 
That is, denoting $f_{jk}$ the color intensity at the $j$th pixel of 
the $k$th image data, the quantum correspondence is given by 
\begin{equation}
    \nonumber
	|{\rm data}(k)\rangle 
	= \frac{1}{\sqrt{N_{P}}} 
	     \sum_{j=0}^{N_P -1} |f_{jk}\rangle_{C} |j\rangle_{P}. 
\end{equation}
This is also a $n_D = n_C + n_P$ qubits state. 
The database state $\ket{{\rm database}}$ is the superposition of the 
above $|{\rm data}(k)\rangle$, as defined in 
Eq.~\eqref{EQ_step1_psi_data_index}. 
Figure~\ref{FIG_fig_setting} depicts an example of such $\ket{{\rm database}}$. 
Note that $\sum_{j=0}^{N_P -1} |f_{jk}\rangle_{C}$ and $\sum_{j=0}^{N_P -1} 
|g_{j}\rangle_{C}$ are normalized. 
Our goal is to identify the index of the elements in $\ket{{\rm database}}$ 
that has the highest similarity with $\ket{{\rm query}}$. 
Note that the probability to hit the index $k$, without the amplitude 
amplification, is calculated as 
\begin{equation}
    \begin{aligned}
	P({\rm index}=k) 
	&= | \mbox{}_I\langle k| \langle 0^{n_D} |\Psi\rangle|^2 \\
	&= \frac{1}{N_I} |\langle {\rm query}|{\rm data}(k)\rangle |^2 \\
	\label{EQ_Px_gf}
	&= \frac{1}{N_I}\left[ \sum_{j=0}^{N_{P}-1} 
	        \frac{\langle g_{j}|f_{jk}\rangle }{N_P} \right]^2 \\
	&\le \frac{1}{N_I}.
	\end{aligned}
\end{equation}

\subsection{Problem formulation}
\label{Section III-B}

\begin{figure}[t!] 
   \includegraphics[width=3.3in]{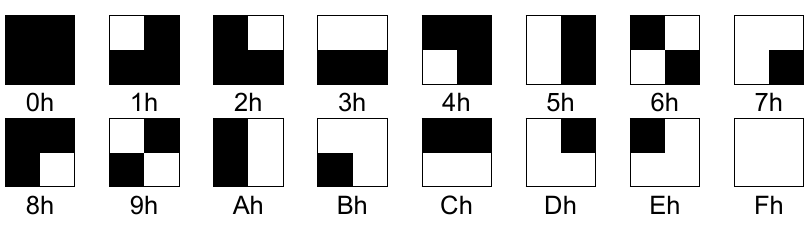}
   \caption{4 pixel-binary data. "h" means the hexadecimal representation.}
   \label{FIG_pix_data}
\end{figure}

The query data is a 4-pixel binary image data, meaning that $n_C=1$ and $n_P=2$. 
(Note that, in this binary image setting, the NEQR is equivalent to the FRQI; 
see Appendix~\ref{SEC_Prelim}.) 
Then, the set of all possible data quantum state is given by 
$\{|0$h$\rangle, |1$h$\rangle, \cdots, |$Fh$\rangle\}$ in the hexadecimal 
representation, as illustrated in Fig.~\ref{FIG_pix_data}. 
Here, the database is chosen as the following subset composed of 8 image 
data (hence, $n_I=3$): 
\begin{align*}
   \{ |{\rm data}(0)\rangle, |&{\rm data}(1)\rangle, |{\rm data}(2)\rangle, |{\rm data}(3)\rangle, \\
   &|{\rm data}(4)\rangle, |{\rm data}(5)\rangle, |{\rm data}(6)\rangle, |{\rm data}(7)\rangle \}  \\
   = \{|0&\mathrm{h}\rangle, |2\mathrm{h}\rangle, |4\mathrm{h}\rangle, |6\mathrm{h}\rangle, |8\mathrm{h}\rangle, |\mathrm{Ah}\rangle, |\mathrm{Ch}\rangle, |\mathrm{Eh}\rangle \}.
\end{align*} 
Therefore, the database quantum state $\ket{{\rm database}}$ is a 6-qubit 
quantum state, composed of $n_C=1$ qubit for the color intensity, $n_P=2$ 
qubits for the pixel position ($X \in \{0,1\}$ and $Y \in \{0,1\}$), and 
$n_I=3$ qubits for the index.

In this demonstration, we consider the following three different settings: 
(i) QASM simulator on Qiskit \cite{Qiskit} is used to simulate AAE and 
the pattern matching part,  
(ii) the optimal parameters $\boldsymbol{\theta_{a}}$ and 
$\boldsymbol{\theta_{b}}$  obtained in the setup (i) are used to run the 
circuit for pattern matching on the superconducting quantum processor 
(ibm\_kawasaki), and 
(iii) the data encoding process is executed via the perfect encoding algorithm 
containing Toffoli gates, rather than AAE; the pattern matching process is 
then operated using the QASM simulator.

We executed AAE to encode the image data as follows. 
The PQC for encoding the database state, $A(\boldsymbol{\theta_a})$, is a 
6-layers hardware efficient ansatz (HEA), and that for the query state, 
$B(\boldsymbol{\theta_b})$, is a 3-layers HEA. 
Each layer is composed of the parameterized single-qubit $Y$-rotational 
gate $R_y(\theta_r)=\exp(-i\theta_r\sigma_y/2)$ and CNOT gate that connect 
adjacent qubits, as shown in Fig.~\ref{circuit}, where $\theta_r$ is the 
$r$-th parameter and $\sigma_y$ is the Pauli $Y$ operator (hence $A(\boldsymbol{\theta_a})$ and $B(\boldsymbol{\theta_b})$ are real matrices). 
We randomly initialized all $\theta_r$, at the beginning of each training.
As for the kernel function, $\kappa(x,y)=\exp(-64(x-y)^{2})$ is used. 
To compute the $r$-th gradient of the loss function, we generate 400 samples 
for each $q_{\theta}^{+}$, $q_{\theta_r}^{-}$, $q_{\theta_r}^{H+}$, and 
$q_{\theta_r}^{H-}$ for training $A(\boldsymbol{\theta_a})$, and 10000 
samples for the case of $B(\boldsymbol{\theta_b})$. 
As the optimizer, Adam \cite{Kingma2014-pa} is used; the learning rate is 0.1 
for the first 100 epochs and 0.01 for the other epochs. 
The number of iterations (i.e., the number of the updates of the parameters) 
for training the PQC is set to 300 for $B(\boldsymbol{\theta_b})$ and 500 for 
$A(\boldsymbol{\theta_a})$.

After the training process of AAE, we run the pattern matching algorithm. 
Both with and without the amplitude amplification via Grover operation, we 
construct the probability distribution $P({\rm index}=k)$, for all 
$k\in \{0,1, \cdots, 7\}$ in Eq.~\eqref{EQ_Px_gf}, to estimate the set of 
indices of the data that has significant similarity to the query. 
The number of samples (or the shot) is chosen depending on a specified 
precision, but it is 512 for all the result displayed in what follows.

\fig{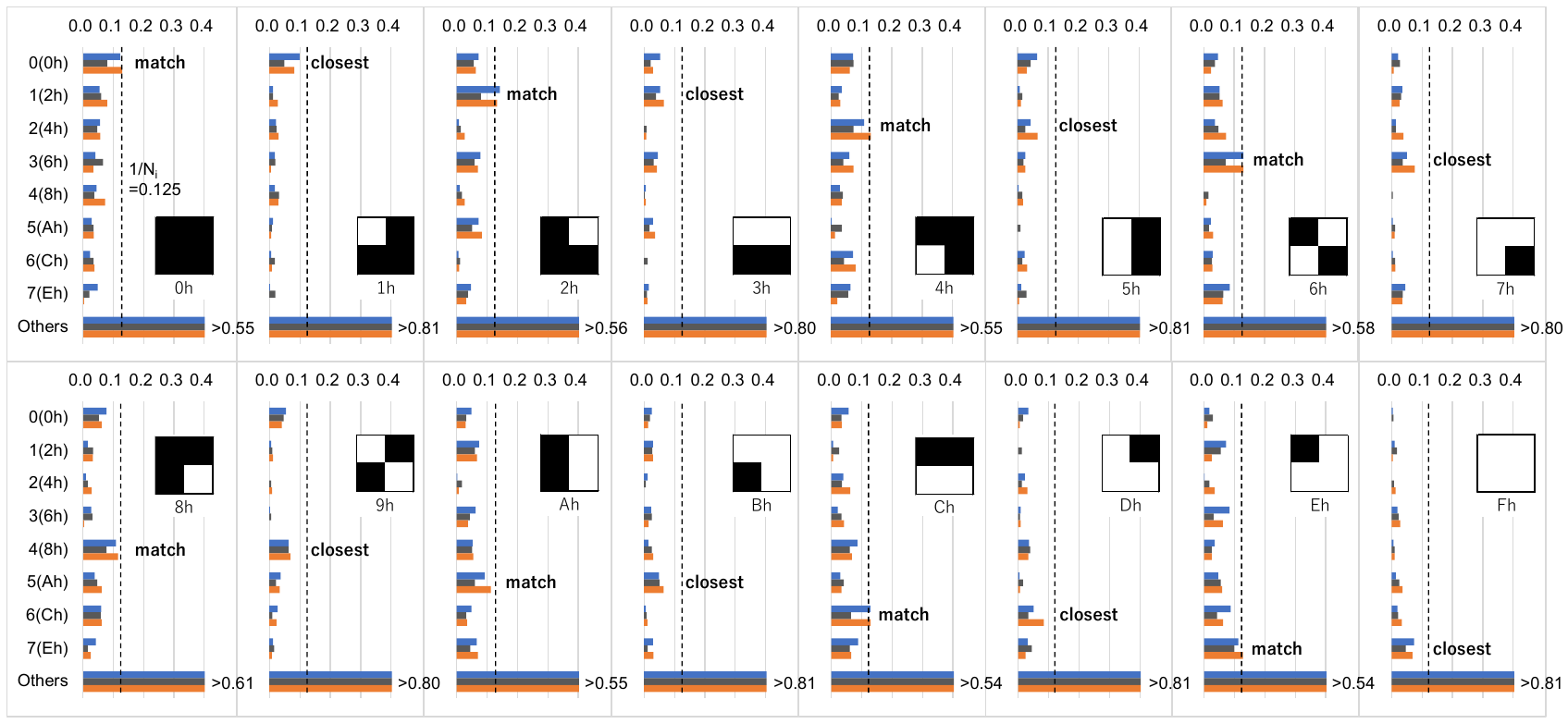}
{Result of the pattern matching test without amplitude amplification. 
The vertical axis is index (data) of the database 
(index 0,1,2,3,4,5,6,7)=(data 0h,2h,4h,6h,8h,Ah,Ch,Eh). 
The bars represent the probability 
$P({\rm index}=k) = | \mbox{}_I\langle k| \langle 0^{n_D} |\Psi\rangle|^2$; 
the blue, gray, and orange bar corresponds to the case (i), (ii), and (iii), 
respectively. 
"Others" in the vertical axis means the failure probability, i.e., the total 
probability of projection of $\ket{\Psi}$ onto the states other than 
$\ket{0}^{\otimes n_D}$. 
"match" represents the index of database component that coincides with the 
query with highest probability. 
"closest" represents the index of database component that is not exactly match 
with but the closest to the query. 
The square images in each figure are the query data.}
{FIG_InvTest_result_0}{width=6.88in}

\fig{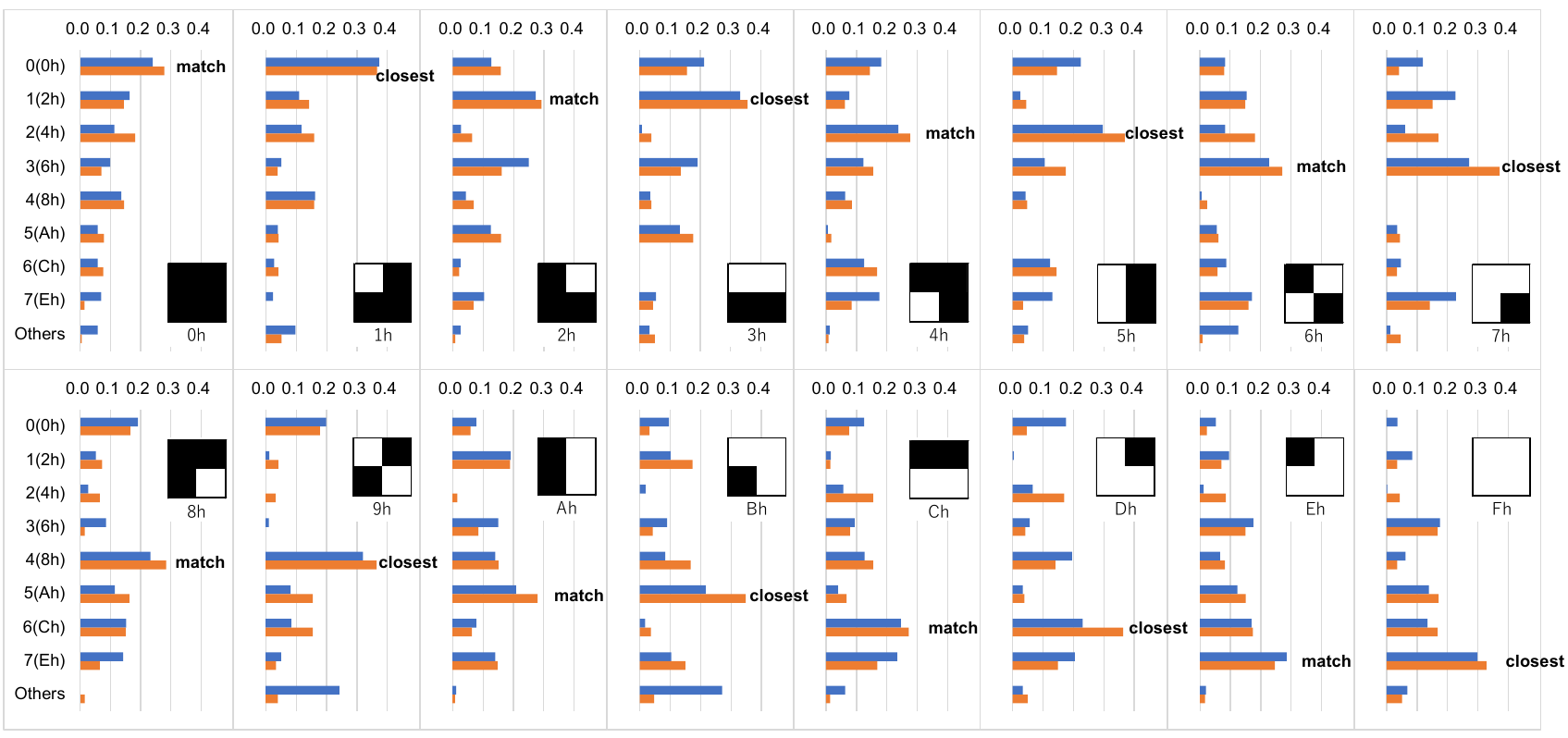}
{Result of the pattern matching test using amplitude amplification, with 
the number of Grover iterations $t=5$. 
"match" represents the index of database component that coincides with the 
query with highest probability. 
"closest" represents the index of database component that is not exactly 
match with but the closest to the query. 
Note that the oracle operator is the same in all cases.}
{FIG_InvTest_result_1}{width=6.88in}

\subsection{Results and discussion}

The probability distribution of the index variable, without amplitude 
amplification (i.e., $t$=0), is shown in Fig.~\ref{FIG_InvTest_result_0}; 
that is, the empirical probability value of 
$P({\rm index}=k) = | \mbox{}_I\langle k| \langle 0^{n_D} |\Psi\rangle|^2$ 
given in Eq.~\eqref{EQ_Px_gf} is calculated by sampling and then horizontally 
displayed, while the indices of the database component, 
(index: 0,1,2,3,4,5,6,7)=(data: 0h,2h,4h,6h,8h,Ah,Ch,Eh), are 
shown in the vertical axis. 
"Others" in the vertical axis means the failure probability, i.e., the total 
probability of projection of $\ket{\Psi}$ onto the states other than 
$\ket{0}^{\otimes n_D}$. 
The blue, gray, and orange bar corresponds to the case (i), (ii), and (iii), 
respectively; see Section \ref{Section III-B} for the meaning of these cases. 
Each subfigure shows the result with respect to the different query data, 
e.g., 0h in the top left.

This result shows that the algorithm works well; for both simulation (i) 
and the experiment (ii), the AAE followed by the inversion test circuit 
produces the outcomes that well agree with the theoretical prediction (iii). 
In particular, the database component obtained with highest probability 
actually coincides with the query, as indicated by "match" in the figure. 
This makes sense because, as shown in Eq.~\eqref{EQ_Px_gf}, the probability 
value with respect to the index corresponds to the overlap, or the fidelity, 
between the database component and the query data, which has the maximum 
value $1/N_{I}$. 
Note that those probability values reflect Hamming distance between two 
images in this case. 
Hence, as a practical use case, a user may set a threshold value of the 
similarity score based on the Hamming distance and apply the result to 
identify some candidate patterns that have the above-threshold overlap 
with the query, e.g., (0h, 2h, 6h, Ah) for the query 2h. 
Moreover, even if the database does not contain exactly the same data as the 
query (hence there is no ``match" index), this idea of posing the candidates 
works; "closest" in the figure represents the index of database component 
that is not exactly match with but the closest to the query.

We now turn our attention to the case with amplitude amplification, under the 
same setting as above. 
We take the number of Grover iterations to be $t=5$. 
The result is shown in Fig.~\ref{FIG_InvTest_result_1}, where the quantities 
in the graph have the same meaning as those in Fig.~\ref{FIG_InvTest_result_0}, 
although in this case we have not conducted the experiment with real quantum 
device. 
Clearly, the amplitude amplification works well, since the failure probability 
of "Others" is drastically reduced and, equivalently, the success probability 
of the post-selection is effectively enhanced. 
As a result, all success probabilities are amplified, while keeping the relative 
ratio due to Eq.~\eqref{EQ_Px_AA_2} showing that all the coefficients $\psi_{x}$ are 
scaled with the same factor of $1/\langle q|\Psi \rangle$. 
Overall, the result obtained with the use of AAE (represented by the blue bars) 
and that with the ideal encoder (orange bars) show good agreement, but some 
deviations are observed. 
This is mainly because of the sampling and encoding errors of AAE. 
In fact, AAE is a machine learning technique to find the approximate encoder, 
so a non-zero error is in general inevitable. 
The impact of such errors on our pattern matching algorithm will be investigated 
in another example, demonstrated in Appendix~\ref{SEC_demo_encoding_err}.

\begin{figure}[t!] 
   \includegraphics[width=3.3in]{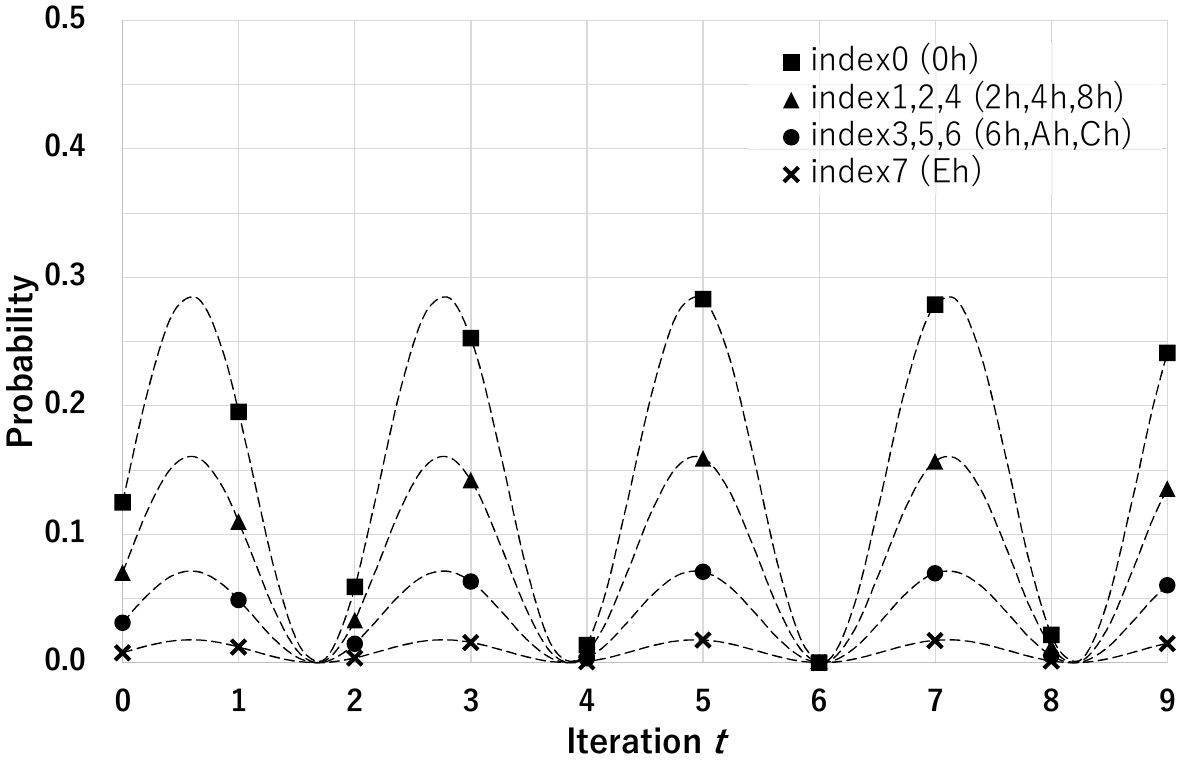}
   \caption{Amplified probability versus the number of iterations. 
   The plots represent the result obtained by the numerical simulation 
   with ideal encoding scheme. 
   The dashed lines represent the analytical expression \eqref{EQ_Px_AA_2}. 
   The database is shown in Fig.~\ref{FIG_InvTest_result_0} and the query data 
   is "0h". }
   \label{FIG_graph_Grover_0h}
\end{figure}

\begin{figure}[t!] 
   \includegraphics[width=3.3in]{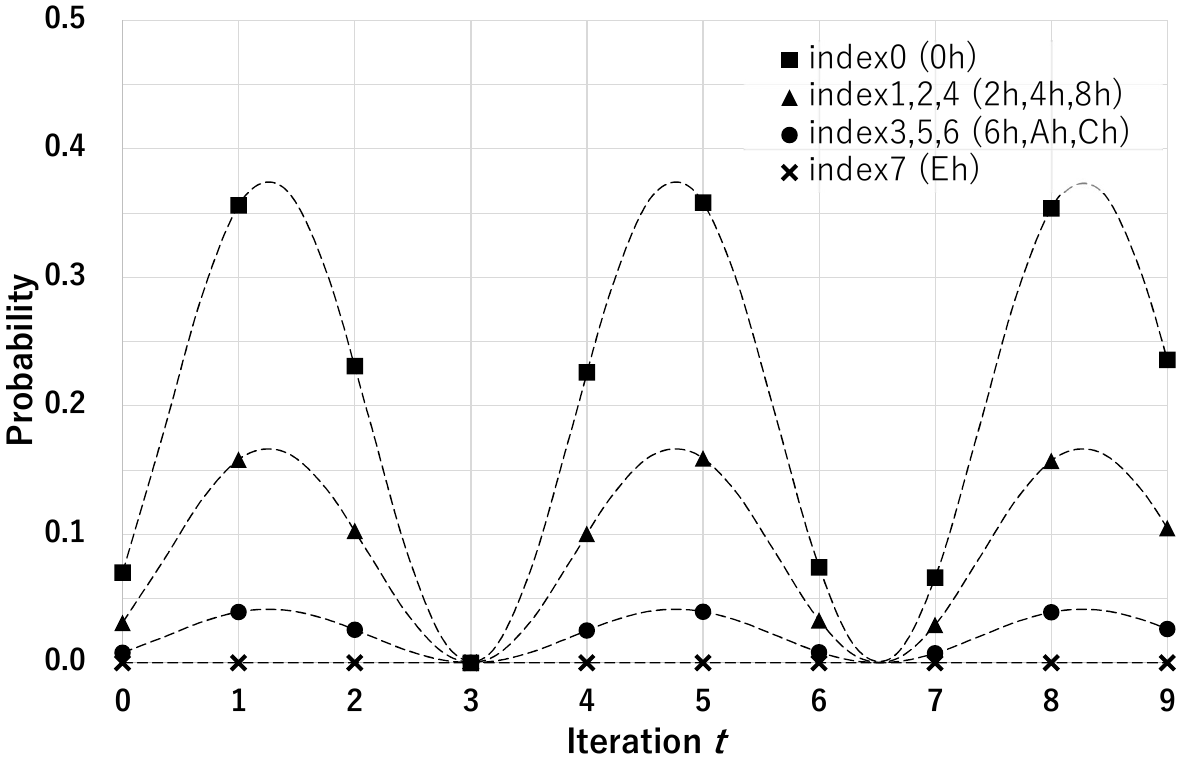}
   \caption{Amplified probability versus the number of iterations. 
   All lines are generated in the same way explained in 
   Fig.~\ref{FIG_graph_Grover_0h}. 
   The query data is here taken as "1h".}
   \label{FIG_graph_Grover_1h}
\end{figure}

Let us now discuss the relation between the number of Grover iterations, $t$, 
and the probability \eqref{EQ_Px_AA_2} to hit the index $k$ of the database 
component, for the case where the query data is 0h or 1h. 
The result is plotted in Fig.~\ref{FIG_graph_Grover_0h} and \ref{FIG_graph_Grover_1h}. 
Clearly, the scaling factor and the frequency of magnification ratio are different 
depending on the query data, reflecting the term $\langle q|\Psi \rangle$. 
As mentioned in Section~\ref{SEC_algo_aa}, to exactly specify the optimal number 
of iteration $t_*$, we need to estimate the value of $\langle q|\Psi \rangle$. 
This task seems to be severe when the frequency of the oscillation 
$\omega = 2 \arcsin{\langle q|\Psi \rangle}$ given in Eq.~\eqref{EQ_omega} is high. 
However, for a practical database where $N_I$ is large, $\omega$ becomes small and 
as a result the probability \eqref{EQ_Px_AA_2} becomes a monotonically increasing 
function with respect to $t$, meaning that a rough choice of $t$ wight work to 
amplify the target probability. 
Another notable point is that 8 indices are categorized to 4 groups. 
These groups reflect Hamming Distance (HD) between the query data and each 
database components, namely HD=0,1,2,3 in Fig.~\ref{FIG_graph_Grover_0h}, and 
HD=1,2,3,4 in Fig.~\ref{FIG_graph_Grover_1h}, from top to bottom. 
This result shows that our algorithm is quite natural in the context of pattern 
matching, although this clear relationship is due to that the data are given by 
the binary images.


\subsection{Depth and the number of multi qubit gate}

Here we discuss the size of circuit for implementing our algorithm. 
First recall that AAE drastically reduces the circuit depth and the number of multi 
qubit gate for implementing a data loading circuit, compared to the conventional 
encoding scheme. 
Actually in the previous subsection, we demonstrated that the 6-qubits database 
quantum state can be well prepared with a 6-layer HEA, which contains 30 CNOT 
gates. 
This is in stark contrast to the conventional exact encoding circuit that requires 
32 6-qubits Toffoli gates, which is decomposed to 128 CNOT gates and 96 auxiliary 
qubits \cite{Maslov2016-mq}. 
One can take a more efficient method \cite{Bergholm2005-pd} that allows us to 
generate an arbitrary $n$ qubit state from $|0\cdots0\rangle$ using $2^n-n-1$ 
CNOT gates (including long-range gates), i.e., at most 57 CNOT gates for the 
case $n=6$. 
However, if we are allowed to use only the nearest-neighbor interaction of 
qubits like current superconducting quantum devices, the number of CNOT gates 
is estimated as follows \cite{Maslov2016-mq}: 
\begin{equation*}
   \frac{10}{3}2^n +2n^2 -12n +
   \begin{cases}
       14/3, ~~ n \;\mathrm{even},\\
       10/3, ~~ n \;\mathrm{odd},
   \end{cases}
\end{equation*}
meaning that at most 218 CNOT gates in the case $n=6$. 
Summarizing, the above exact encoding methods need an exponential number of 
CNOT gates, which is quite challenging. 

In contrast, with PQC, the depth and the number of multi qubit gates seem not 
to increase so fast, i.e., $O$(poly($n$)) in depth and $O$($n$) of multi qubit 
gates for $n$-qubits systems, although a PQC does not guarantee a very precise 
data loading. 
To validate the use of PQC, several theoretical and numerical investigation 
have been conducted. 
For instance, the power of expressibility of various type of PQC is under 
actively studied \cite{Sim2019-zf,Cerezo2020-ue,Holmes2021-ug,Caro2021-jt}; 
actually we find some hardware-efficient circuit ansatz with higher expressibility, 
that are yet relatively easier to implement. 
Those methods may be applicable to the data loading circuit.

Lastly, in our demonstration we naively implemented the diffusion operator $G_d$, 
but there are several implementation schemes which reduce the number of elementary 
gates and the circuit depth to compose the diffusion operator 
\cite{Grover2002-zf,Liu2021-fk,Brianski2021-vd}, which were recently demonstrated  
\cite{Hlembotskyi2020-uy,Gwinner2020-me,Zhang2021-fx}. 
Application of those method will enable us to demonstrate the amplitude amplification 
process executed on a real quantum device.


\section{Conclusions}
\label{SEC_concl}

In this paper, we proposed a method for approximately executing the database 
search or more broadly the pattern matching algorithm, i.e., the algorithm 
that Grover originally considered as a practical application of quantum 
computation. 
The key idea is to implement the data loading process on a shallow parametrized 
quantum circuit and the pattern matching function on the inversion-test based 
circuit, followed by the amplitude amplification operation that can be constructed 
without using the target index. 
The data-loading circuit needs much less multi-qubit entangling gates than the 
conventional one; introducing the recent technique to implement the diffusion 
operator with small blocks \cite{Brianski2021-vd} may reduce the difficulty to 
implement the amplitude amplification part as well. 
Our proposed framework will be then totally beneficial for NISQ but also for 
FTQC devices. 
The demonstration of algorithm in the problem of image pattern matching, with 
both numerical simulations and partially a real quantum device, contributes 
to the field of quantum image processing.

Lastly we discuss the possible quantum advantage of the proposed method, where 
of course the computational overhead in the variational part of AAE should be 
carefully taken into account. 
First, note that the computational complexity in the entire algorithm needs not 
include the cost for preparing the database quantum state as in the classical 
case. 
Therefore the core of complexity lies in the overhead to prepare the query 
quantum state; 
if the number of repetition of the AAE circuit $B(\boldsymbol{\theta})$ becomes 
dominant over the number of Grover operations, the quantum advantage will vanish. 
Thus, it is important to limit each data vector to a low-dimensional one such 
as that of a telephone number or a low-resolution image used for object recognition, 
which can be represented with e.g., up to 6 qubits. 
In this case we expect that AAE would take roughly hundreds of iterations of 
$B(\boldsymbol{\theta})$ to encode the query data with enough precision, or even 
the quantum circuit for exact encoding could be implementable nearly perfectly. 
The point is that, in this scenario, the complexity for preparing the quantum 
query data state does not scale with the number of database, $N_I$, which is 
usually quite big; 
the size of phone book is much bigger than the dimension of each data of phone 
number! 
Hence our view is that, when $N_D \ll N_I$ and a very accurate realization of 
the database state is not required, the proposed algorithm for search or pattern 
match may have a practical quantum advantage both in memory and query complexities. 
Surely the hardness of near-perfect implementation of Grover operator remains, 
which yet might be attacked via several recent proposals 
\cite{Liu2021-fk,Brianski2021-vd} together with possible circumvention via 
exploiting the specific structure of $A$ and $B$. 
These important problems, along with customizing our framework for other 
applications, will be presented elsewhere.

\section*{Acknowledgments}
This work was supported by MEXT Quantum Leap Flagship Program Grant Number 
JPMXS0118067285 and JPMXS0120319794.

\bibliographystyle{apsrev4-2}
\bibliography{main}


\appendix       

\section{Grover algorithm}
\label{SEC_grover_dq}

Here we show the Grover algorithm, in a generalized form where the target 
state (the query state in our language) is not limited to an eigenstate of 
the initial state. 
See e.g., \cite{Ezhov2000-sf} for a more detailed description. 

We first define an initial state $|\Psi \rangle$ as follows:
\begin{equation}
    \begin{aligned}
	\label{EQ_ape_def_Psi}
	|\Psi \rangle &\equiv \sin \theta | q \rangle + \cos \theta | q^{\perp} \rangle, \\
	\ket{q} &\equiv \frac{1}{N} P \ket{\Psi}, \\
	\ket{q ^{\perp}} &\equiv \frac{1}{N'} (\mathbb{1} - P) \ket{\Psi}. 
	\end{aligned}
\end{equation}
That is, $\ket{\Psi}$ is decomposed to a given target state $|q \rangle$ and its 
orthogonal state $|q^{\perp} \rangle$, where $N = \sqrt{\bra{\Psi} P \ket{\Psi}}$ 
and $N' = \sqrt{\bra{\Psi} (\mathbb{1} -P) \ket{\Psi}}$ are the normalization constants. 
$P$ is a projection matrix, which defines the component(s) of $\ket{\Psi}$ to be searched. 
Also note that $\sin\theta=\pro{q}{\Psi}$. 
The goal is to amplify the coefficient of the target state, i.e., $\sin\theta$, via acting the 
Grover operator 
\begin{equation*}
	G \equiv G_{d} G_{o},
\end{equation*}
where $G_{o}$ and $G_{d}$ are the oracle operator and the diffusion operator 
defined as
\begin{equation*}
    \begin{aligned}
	\label{EQ_ape_def_G}
	G_{o} &\equiv \mathbb{1} -2 P, \\
	G_{d} &\equiv 2|\Psi \rangle \langle \Psi | -\mathbb{1}.
	\end{aligned}
\end{equation*}
Actually, by combining the above equations, we have 
\begin{equation*}
    \label{EQ_ape_G_theta}
    G \begin{bmatrix} |q \rangle \\ |q^{\perp} \rangle \end{bmatrix} = 
    \begin{bmatrix} \cos(2 \theta) & -\sin(2 \theta) \\ \sin(2 \theta) & \cos(2 \theta) \end{bmatrix}
    \begin{bmatrix} |q \rangle \\ |q^{\perp} \rangle \end{bmatrix}.
\end{equation*}
This represents that $G$ induces a rotation by $2\theta$ in the space spanned by 
$|q \rangle$ and $|q ^{\perp} \rangle$. 
As a result, we found that $|\Psi ^{(t)} \rangle \equiv G^{t} |\Psi \rangle$ is 
given by 
\begin{equation}
    \label{EQ_ape_GMPsi}
    |\Psi ^{(t)} \rangle 
       = \sin((2t+1)\theta) |q \rangle + \cos((2t+1)\theta) |q^{\perp} \rangle. 
\end{equation}
Hence certainly the amplitude of $\ket{q}$ is amplified, by appropriately choosing 
the number of operations.

Next, let us express $\ket{\Psi}$ and $\ket{q}$ in terms of the basis states $\{ \ket{x} \}$ 
as follows: 
\begin{align}
	\label{EQ_ape_def_ax}
	|\Psi \rangle &\equiv \sum_{x} \psi_x |x\rangle,  \\
	\label{EQ_ape_def_bx}
	|q\rangle &\equiv \sum_{x} q_x |x\rangle. 
\end{align}
For simplicity, we assume $\psi_x$ and $q_x$ are real numbers for all $x$. 
Correspondingly, $|\Psi^{(t)} \rangle$ can also be expressed using the basis 
states $\{ \ket{x} \}$: 
\begin{equation*}
	\label{EQ_ape_def_amx}
	|\Psi^{(t)} \rangle \equiv \sum_{x} \psi_{x}^{(t)} |x\rangle. 
\end{equation*}
We now derive the explicit representation of $\psi_{x}^{(t)}$, to see how each coefficient $\psi_x$ 
of $\ket{\Psi}$ is amplified. 
From Eqs.~\eqref{EQ_ape_def_Psi}, \eqref{EQ_ape_def_ax}, and \eqref{EQ_ape_def_bx}, 
we have
\begin{equation}
    \label{EQ_ape_def_bpx}
    \begin{aligned}
	|q ^{\perp}\rangle &= 
	 \frac{1}{\cos{\theta}} |\Psi \rangle  
	-\frac{\sin{\theta}}{\cos{\theta}} | q \rangle \\
	&= \sum_{x} \biggl[ \frac{\psi_x}{\cos{\theta}} -\frac{\sin{\theta}}{\cos{\theta}} q_x \biggr] |x \rangle \\
	&\equiv \sum_{x} q^{\perp}_{x} |x \rangle.
	\end{aligned}
\end{equation}
Then, by substituting Eqs.~\eqref{EQ_ape_def_bx} and \eqref{EQ_ape_def_bpx} 
into Eq.~\eqref{EQ_ape_GMPsi}, we have
\begin{equation*}
    \begin{aligned}
	|\Psi^{(t)} \rangle  
	&= \sum_{x} \biggl[ \sin((2t+1)\theta) q_x +\cos((2t+1)\theta) q^{\perp}_x \biggr] |x \rangle.
	\end{aligned}
\end{equation*}
Therefore,
\begin{equation}
\label{appendix a_x^t}
    \begin{aligned}
	\psi^{(t)}_x &=  \sin((2t+1)\theta) q_x +\cos((2t+1)\theta) 
	     \frac{\psi_x - q_x\sin{\theta}}{\cos{\theta}} \\
	&= \frac{1}{\cos{\theta}} \biggl[ q_x \sin(2t\theta) +\psi_x \cos(2t\theta +\theta) \biggr] \\
	&= \frac{1}{\cos{\theta}} 
	    \biggl[ (q_x -\psi_x \sin{\theta}) \sin(2t\theta) +\psi_x \cos{\theta} \cos(2t\theta)\biggr] \\
	&= \sqrt{\frac{\psi_x ^2 -2\langle q|\Psi \rangle \psi_x q_x +q_x ^2}{1-\langle q|\Psi \rangle ^2}} 
	        \sin(2t\theta +\phi_x), 
	\end{aligned}
\end{equation}
where
\begin{equation*}
    \begin{aligned}
	\theta &= \arcsin \langle q|\Psi \rangle, \\
	\phi_x &= \arccos \left( \frac{q_x -\langle q|\Psi \rangle \psi_x}
	                        {\sqrt{\psi_x ^2 -2\langle q|\Psi \rangle \psi_x q_x +q_x ^2}} \right).
	\end{aligned}
\end{equation*}
Hence the phase $\phi_x$ determines the amplification gain as a function of $x$. 
Note that, in the original Grover algorithm, $\ket{\Psi}$ is the superposition of equiprobable 
basis states, i.e., $\psi_{0}= \cdots = \psi_{N-1}= 1/\sqrt{N}$, and the target state is one of them, 
meaning that $P=\ket{y}\bra{y}$ and accordingly $\ket{q}=\ket{y}$ and $q_x=\delta_{x,y}$. 
In this case, $\sin\theta = \pro{q}{\Psi} = 1/\sqrt{N}$, which leads to 
\begin{equation*}
     \psi_y^{(t)} = \sin(2t\theta + \phi_y), ~~
     \phi_y = \arccos\sqrt{1-1/N} = \theta, 
\end{equation*}
and for $x\neq y$ 
\begin{equation*}
     \psi_x^{(t)} = \frac{\sin(2t\theta + \phi_x)}{\sqrt{N-1}}, ~~
     \phi_x = \arccos\frac{1}{\sqrt{N}} = \theta + \frac{\pi}{2}. 
\end{equation*}
This coincides with the formula given in the original Grover algorithm 
\cite{Boyer1998-ez}.

In our context, the initial state is $\ket{\Psi}=\sum_{x}\psi_x \ket{x}$ with $\{ \ket{x} \}$ 
the set of basis states of ${\cal H}_D\otimes {\cal H}_I$. 
The target state $\ket{q}$ is characterized by the projection 
$P=\ket{0^{n_D}}\bra{0^{n_D}}\otimes \mathbb{1}_I$; 
that is, $\ket{q}$ is the state having the query data state $\ket{{\rm query}'}=\ket{0}^{\otimes n_D}$ 
for any index component. 
Then we have 
\begin{equation*}
     \ket{q} = \frac{1}{N} \sum_x \psi_x P \ket{x}
                = \frac{1}{N} \sum_{x\in{\cal C}} \psi_x \ket{x}, 
\end{equation*}
where ${\cal C}$ is the set of numbers defined as 
\begin{equation*}
     {\cal C} = \{ x ~|~ \bra{x} (\ket{0^{n_D}}\bra{0^{n_D}}\otimes \mathbb{1}_I) \ket{x} \neq 0 \}. 
\end{equation*}
Note that $N=\pro{q}{\Psi} 
= \left[ \bra{\Psi} (|0^{n_D}\rangle \langle 0^{n_D}| \otimes \mathbb{1}_I ) \ket{\Psi} \right]^{1/2}$. 
As a result, 
\begin{equation*}
	q_x =
	\begin{cases}
	    \psi_{x}/\langle q | \Psi\rangle  & \mathrm{if} \; x \in {\cal C}, \\
	    0   & \mathrm{if} \; x \notin {\cal C}. 
	    \end{cases}
\end{equation*}
Then the state after $t$ Grover operations, $|\Psi^{(t)}\rangle 
= G^t |\Psi \rangle = \sum_{x} \psi_{x}^{(t)} |x\rangle$, can be specified as follows; 
that is, from Eq.~\eqref{appendix a_x^t}, we find that 
$\psi_{x}^{(t)} = A_{x} \sin(\omega t +\delta_{x})$ where $\omega=2\theta$, $A_{x}$, and 
$\delta_{x}$ are given by Eqs.~\eqref{EQ_omega}, \eqref{EQ_Ax}, and \eqref{EQ_deltax}, respectivley. 
Also the hitting probability $P_{x}^{(t)} = \{ \psi_{x}^{(t)} \} ^2$ can also be readily calculated and 
given by Eq.~\eqref{EQ_Px_AA_2}.

\section{Quantum Image representations}
\label{SEC_Prelim}

Various quantum image representation frameworks have been proposed 
\cite{Wang2021-fn}. 
They are categorized to the amplitude-encoding-based format and the 
basis-encoding-based format. 
The basic idea of the former is summarized by Flexible Representation 
for Quantum Images (FRQI), and the latter is summarized by Novel Enhanced 
Quantum Representation (NEQR). 
In this section, these two representations are reviewed; note that both 
representations can be taken in our proposed algorithm.

\subsection{Flexible Representation for Quantum Images (FRQI)}

In general, quantum image representation is a method for expressing the pixel 
positions and the corresponding color intensities, as a quantum superposition 
state. 
In the FRQI format \cite{Le2011-wq}, a grayscale image on $N_p = 2^n \times 2^n$ 
pixels is encoded in a quantum state as follows; 
\begin{equation}
    \label{EQ_formula_FRQI}
    \begin{split}
    |I_{FRQI}\rangle &= \frac{1}{\sqrt{N_p}} \sum_{z=0}^{N_p -1} \left(\cos\theta_{z} |0\rangle +\sin\theta_{z} |1\rangle\right) |z\rangle  \\
    &= \frac{1}{\sqrt{N_p}} \sum_{z=0}^{N_p -1} |f_{z}\rangle |z\rangle, 
    \end{split} 
\end{equation}
where $|0\rangle$ and $|1\rangle$ are the computational basis states of a 
single qubit, while $\{|z\rangle \}$ is the computational basis states of 
$\log N_p=2n$ qubits representing the coordinate of the corresponding pixel. 
$\boldsymbol{\theta}=(\theta_0,\theta_1,\cdots,\theta_{2^{2n}-1}), \theta_i \in [0, \pi/2]$ 
is the vector of angles encoding the colors; 
that is, $|f_z\rangle = \cos\theta_{z} |0\rangle +\sin\theta_{z} |1\rangle$ 
is a qubit on which the color at the pixel coordinate $z$ is encoded. 
That is, a gray scale $2^n\times2^n$ pixel-image is represented by the 
$2n+1$ qubits state \eqref{EQ_formula_FRQI}.

The point of FRQI is that it needs only the minimum number of qubit in which one 
can encode the absolute value of color information with arbitrary precision. 
This is advantageous to amplitude encoding frameworks such as \cite{Yao2017-nr}, 
where the color information is directly encoded into the amplitude of state 
vectors representing the pixel position. 
Actually, the scheme \cite{Yao2017-nr} cannot represent the absolute color value 
but only the relative value, in contrast to FRQI and NEQR; 
this induces an undesirable change of amplitude information depending on the other 
pixel's one, which happens even if the absolute intensities are the same, e.g., 
in the case where only one pixel has maximum intensity and others are zero or the 
case where two pixels have the maximum intensity.

On the other hand, FRQI has the following disadvantage; that is, exponential number 
of measurement is needed to restore the exact image since one has to identify all 
the state amplitudes. 
However, if we are interested only in the intermediate process rather than retrieving 
the complete image information, this measurement issue is not crucial, as pointed out 
in \cite{Ruan2021-qs}. 
Actually, our pattern matching problem does not need such an exact information 
retrieval, and user can adjust the sampling cost only depending on the required 
accuracy.

Finally, in this FRQI representation, the probability distribution with respect to 
the index variable, \eqref{EQ_Px_gf}, is explicitly given by 
\begin{align}
	\label{EQ_step3_prob_frqi}
	\begin{split}
	& P_{FRQI}({\rm index}=x) 
	   = \frac{1}{N_I} \left( \frac{N_{p}^{FRQI}(x)}{N_p} \right)^2, \\
	& N_{p}^{FRQI}(x) 
	  = \sum_{j=0}^{N_{p}-1} 
	     \Big(\cos\theta_{j} \cos\theta_{jx} +\sin\theta_{j} \sin\theta_{jx}\Big),
	\end{split} 
\end{align}
where $\theta_{j}$ and $\theta_{jx}$ correspond to the color intensity of $j$th pixel in the query data 
and the color intensity of $j$th pixel of $x$th image in the database, respectively.

\subsection{ Novel Enhanced Quantum Representation (NEQR)}

NEQR proposed in \cite{Zhang2013-ig} is used for encoding a digital image, meaning 
that the color intensity is represented by a binary code. 
In this scheme, the color information represented by the discrete variable taking 
the $2^q$ values, can be stored in a $q$-qubit system. 
If the image is created on $N_p =2^n \times 2^n$ pixels, it is encoded into the 
quantum state of the form
\begin{equation}
    \label{EQ_formula_NEQR}
    \begin{split}
    |I_{NEQR}\rangle &= \frac{1}{\sqrt{N_p}} \sum_{z=0}^{N_p -1} \otimes_{i=0}^{q-1} |c_{z}^{i}\rangle |z\rangle  \\
    &= \frac{1}{\sqrt{N_p}} \sum_{z=0}^{N_p -1} |f(z)\rangle |z\rangle, 
    \end{split} 
\end{equation}
where $c_{z}^{i}\in \{0,1\}$ is a bit information and 
$f(z)=\otimes_{i=0}^{q-1} c_{z}^{i}\in [ 0, 2^{q}-1 ]$ is a bit string representing 
the pixel color information at the coordinate $z$. 
Clearly, the total number of qubit depends on the level of quantization of the color 
information, and fine encoding needs much more qubits than FRQI. 
On the other hand, one can restore an $N_p$-pixel quantum image exactly with $O$($N_p$) 
measurement, thanks to the orthogonality of the states. 
In this NEQR representation, the probability distribution with respect to 
the index variable, \eqref{EQ_Px_gf}, is given by 
\begin{align}
	\label{EQ_step3_prob_neqr}
	\begin{split}
	& P_{NEQR}({\rm index}=x) 
	   = \frac{1}{N_I} \left( \frac{N_{p}^{NEQR}(x)}{N_p} \right)^2, \\
	& N_{p}^{NEQR}(x) 
	   = \sum_{j=0}^{N_{p}-1} \otimes_{i=0}^{q-1}  \langle c_j^i|c_{jx}^i\rangle,
	\end{split} 
\end{align}
where $c^{i}_{j}$ and $c^{i}_{jx}$ represent $i$th bit of the color intensity bit string of $j$th pixel in the query data 
and $i$th bit of the color intensity bit string of $j$th pixel of $x$th image in the database, respectively.


\subsection{Impact of the encoding error for the case of handwritten digit data}
\label{SEC_demo_encoding_err}

Here we show the result of analysis on the impact of encoding error, depending 
on the encoding schemes; 
that is, we will display the two probability distributions 
\eqref{EQ_step3_prob_frqi} and \eqref{EQ_step3_prob_neqr} under some errors. 
We use ’Handwritten Digits’ dataset \cite{digit_dataset} composed of 10 
handwritten digits from the number 0 to 9; each data is a $8 \times 8$ pixels image 
with 16 level color assigned by the amplitude variable $u_x \in [0, 1]$, where 
$x$ denotes the position index of the image. 
We suppose that the amplitude $u_x$ is affected by the noise $\epsilon_x$ 
generated from the normal distribution 
$N(0,\sigma), \sigma = \max(\{u_{x}\}) \sigma_{0} \}$. 
Then we make the normalized amplitude vector $\{u_{x} +\epsilon_{x} | x\in$ all 
basis$\}$ divided by $\sqrt{\sum_{x} (u_{x} +\epsilon_{x})^2}$. 
The above procedure are conducted for all the database component and a query data. 
Then we calculate the probability distributions \eqref{EQ_step3_prob_frqi} and 
\eqref{EQ_step3_prob_neqr} for several noise deviation 
$\sigma_{0} \in \{0.05, 0.1, 0.3, 0.5\}$ to see the impact of noise. 
Note that we consider the case without amplitude amplification. 
In the case with amplitude amplification, the amplitudes are magnified 
the factor of $1/\langle q|\Psi\rangle$ while keeping the relative amplitude.

\fig{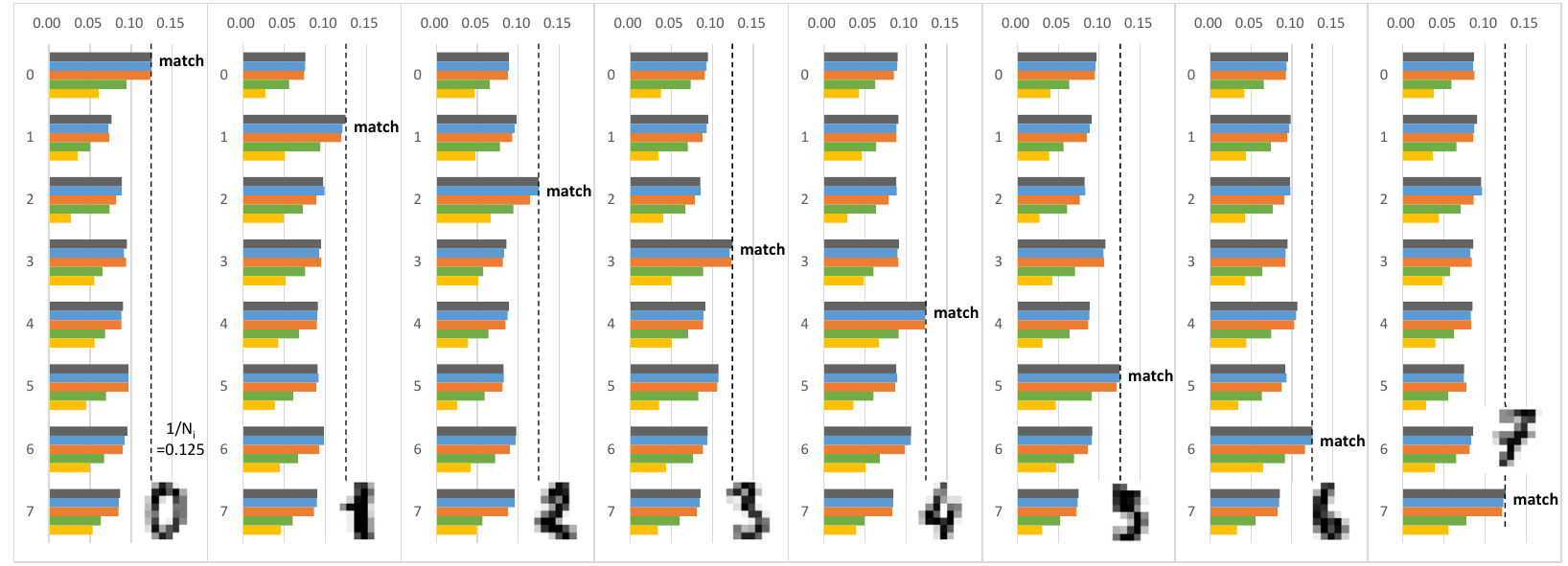}
{
Result of the pattern matching test when using FRQI. 
The vertical axis is index (data) of the database 
(index 0,1,2,3,4,5,6,7)=(digit 0,1,2,3,4,5,6,7). 
The horizontal axis is the probability $P_{FRQI}({\rm index}=x)$. 
"match" represents the index of database component that coincides with the 
query with highest probability. 
The gray bar is the case of ideal encoding. 
The blue, orange, green, and yellow bars are the cases where the encoding noise 
are 5\%, 10\%, 30\% and 50\% (corresponding fidelity values are 0.99, 0.98, 0.86, 
and 0.72, respectively). 
The single data is represented using 7 qubits (1 qubit for the color intensity 
and the others for position), and the database is represented using 10 qubits 
since it has 8 indices.}
{FIG_noise_FRQI}{width=6.9in}

\fig{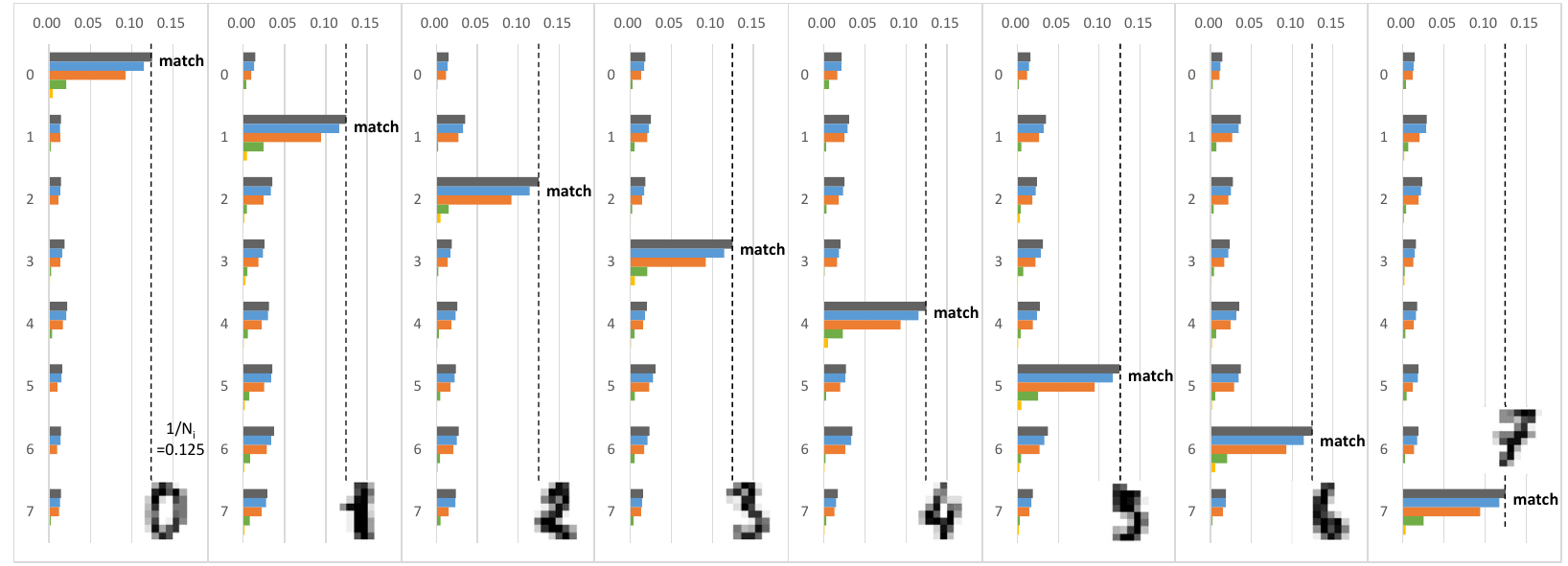}
{
Result of the pattern matching test when using NEQR. 
The vertical axis is index (data) of the database 
(index 0,1,2,3,4,5,6,7)=(digit 0,1,2,3,4,5,6,7). 
The horizontal axis is the probability $P_{NEQR}({\rm index}=x)$. 
The gray bar is the case of ideal encoding. 
The blue, orange, green, and yellow bars are the cases where the encoding noise 
are 5\%, 10\%, 30\% and 50\% (corresponding fidelity values are 0.96, 0.86, 0.41, 
and 0.20, respectively). 
The single data is represented using 10 qubits (4 qubit for the color intensity 
(16 levels) and the others for position), and the database is represented using 
13 qubits since it has 8 indices. 
}
{FIG_noise_NEQR}{width=6.9in}

Results are shown in Figs.~\ref{FIG_noise_FRQI} and \ref{FIG_noise_NEQR}. 
Both of FRQI and NEQR work appropriately, giving the correct indices with 
highest score. 
In terms of distinguishability, NEQR seems better, because of the clear difference 
between the matched state and the others. 
However, this also means that we cannot see well the similarity degree between 
each components. 
In other words, FRQI is appropriate for fuzzy matching, and NEQR for exact 
matching.
For example, we can see in Fig.~\ref{FIG_noise_FRQI} that "3" and "5" are similar, 
since those scores are higher than others (Index 5 in Query 3 and Index 3 in Query 5)
Thus one should choose the data format depending on the purpose.

In terms of error tolerance, FRQI is better. 
When the noise magnitude is less than 10\%, it does not give a large affection 
on the similarity score in FRQI. 
On the other hand in NEQR, some non-negligible impact is observed even with 5\% 
error, and moreover, the score drastically becomes small with 30\% error. 
These behavior is explained by their representation formats. 
In FRQI, the color intensity is expressed by a single qubit, 
$\cos\theta |0\rangle + \sin\theta |1\rangle$, which is independent to the 
color-level precision, and most sub-states have non-zero amplitude with ordinary 
images (highlight or shadow clipping are rare). 
If the encoding error occurs randomly on each states, the effect can be 
compensated each other. 
However, in NEQR, the data is encoded in a sparse space, or equivalently the 
sub-state is expressed by a one-hot vector; e.g., for the case of 4 qubit with 
16 color levels, only one sub-state has non-zero amplitude and the other 15 
sub-states have zero amplitude. 
Since the similarity score directly relates to the inner-product of the noised 
one-hot vectors, it is susceptible to the encoding error.

\end{document}